\documentclass[twocol]{ametsocV5_arxiv}

\usepackage{amsmath,amsfonts,amssymb,bm}
\usepackage{mathptmx}%{times}
\usepackage{newtxtext}
\usepackage{cjhebrew}
\usepackage{newtxmath}
\usepackage{multirow}

\usepackage{bbm}
\usepackage{graphicx}
\usepackage{soul}

\newcommand{\pr}{\mathbb{P}}
\newcommand{\ex}{\mathbb{E}}
\newcommand{\gcal}{\mathcal{G}}
\newcommand{\tcal}{\mathcal{T}}
\newcommand{\ical}{\mathcal{I}}

\newcommand{\lcal}{\mathcal{L}}

\newcommand{\acal}{\mathcal{A}}
\newcommand{\bigo}{\mathcal{O}}
\newcommand{\ind}{\mathbbm{1}}

\newcommand{\norm}[1]{\lVert#1\rVert}

\newcommand{\ang}[1]{\langle#1\rangle}
\newcommand{\tr}{^\top}
\newcommand{\cm}{^\textrm{c}}
\newcommand{\inv}{^{-1}}

\newcommand{\vx}{\mathbf{x}}
\newcommand{\bx}{\mathbf{X}}
\newcommand{\vy}{\mathbf{y}}

\newcommand{\vb}{\mathbf{b}}
\newcommand{\va}{\mathbf{a}}

\newcommand{\khat}{\mathbf{k}}

\newcommand{\jhat}{\mathbf{j}}

\newcommand{\ov}{\overline}
\newcommand{\re}{\text{Re}}
\newcommand{\im}{\text{Im}}
\newcommand{\fn}{\frac{f_0^2}{N^2}}
\newcommand{\der}[2]{\frac{d{#1}}{ d{#2}}}
\newcommand{\pder}[2]{\frac{\partial{#1}}{\partial{#2}}}
\newcommand{\pdersq}[2]{\frac{\partial^2{#1}}{\partial{#2}^2}}

\newcommand{\eps}{\varepsilon}
\newcommand{\bmt}{\begin{bmatrix}}
\newcommand{\emt}{\end{bmatrix}}

\newcommand{\str}{^*}
\newcommand{\s}[1]{^{(#1)}} %s for superscript (but with parentheses this time)

\newcommand{\R}{\mathbb{R}}

\newcommand{\qp}{q^+}
\newcommand{\qm}{q^-}

\newcommand{\lt}{\eta^+}
\newcommand{\dam}{\Gamma}

\title{Learning forecasts of rare stratospheric transitions from short simulations}

\authors{Justin Finkel\correspondingauthor{Justin Finkel, jfinkel@uchicago.edu}}
\affiliation{Committee on Computational and Applied Mathematics, University of Chicago}

\extraauthor{Robert J. Webber}
\extraaffil{Department of Computing and Mathematical Sciences, California Institute of Technology}

\extraauthor{Edwin P. Gerber}
\extraaffil{Courant Institute of Mathematical Sciences, New York University}

\extraauthor{Dorian S. Abbot}
\extraaffil{Department of the Geophysical Sciences, University of Chicago}

\extraauthor{Jonathan Weare}
\extraaffil{Courant Institute of Mathematical Sciences, New York University}

\abstract
{Rare events arising in nonlinear atmospheric dynamics remain hard to predict and attribute. We address the problem of forecasting rare events in a prototypical example, Sudden Stratospheric Warmings (SSWs). Approximately once every other winter, the boreal stratospheric polar vortex rapidly breaks down, shifting midlatitude surface weather patterns for months. We focus on two key quantities of interest: the probability of an SSW occurring, and the expected lead time if it does occur, as functions of initial condition. These \emph{optimal forecasts} concretely measure the event's progress. Direct numerical simulation can estimate them in principle, but is prohibitively expensive in practice: each rare event requires a long integration to observe, and the cost of each integration grows with model complexity.  We describe an alternative approach using integrations that are \emph{short} compared to the timescale of the warming event. We compute the probability and lead time efficiently by solving equations involving the transition operator, which encodes all information about the dynamics. We relate these optimal forecasts to a small number of interpretable physical variables, suggesting optimal measurements for forecasting. We illustrate the methodology on a prototype SSW model developed by Holton and Mass (1976) and modified by stochastic forcing. While highly idealized, this model captures the essential nonlinear dynamics of SSWs and exhibits the key forecasting challenge: the dramatic separation in timescales between a single event and the return time between successive events. Our methodology is designed to fully exploit high-dimensional data from models and observations, and has the potential to identify detailed predictors of many complex rare events in meteorology.  
}

\begin{document}

%% Necessary! 
\maketitle

%%%%%%%%%%%%%%%%%%%%%%%%%%%%%%%%%%%%%%%%%%%%%%%%%%%%%%%%%%%%%%%%%%%%%
% SIGNIFICANCE STATEMENT/CAPSULE SUMMARY
%%%%%%%%%%%%%%%%%%%%%%%%%%%%%%%%%%%%%%%%%%%%%%%%%%%%%%%%%%%%%%%%%%%%%
%
% If you are including an optional significance statement for a journal article or a required capsule summary for BAMS 
% (see www.ametsoc.org/ams/index.cfm/publications/authors/journal-and-bams-authors/formatting-and-manuscript-components for details), 
% please apply the necessary command as shown below:
%
% \statement
% Significance statement here.
%
% \capsule
% Capsule summary here.

%%%%%%%%%%%%%%%%%%%%%%%%%%%%%%%%%%%%%%%%%%%%%%%%%%%%%%%%%%%%%%%%%%%%%
% MAIN BODY OF PAPER
%%%%%%%%%%%%%%%%%%%%%%%%%%%%%%%%%%%%%%%%%%%%%%%%%%%%%%%%%%%%%%%%%%%%%
%

\section{Introduction}
As computing power increases and weather models grow more intricate and  capable of generating a vast wealth of realistic data, the goal of extreme weather event prediction appears less distant \citep{Vitart2018}. To take full advantage of the increased computing power, we must develop new approaches to efficiently manage and parse the data we generate (or observe) to derive physically interpretable, actionable insights. Extreme weather events are worthy targets for simulation owing to their destructive potential to life and property. Rare events have attracted significant simulation efforts recently, including hurricanes \citep[e.g.,][]{Zhang2009,webber,Plotkin2019},  heat waves \citep[e.g.,][]{Ragone24}, rogue waves \citep[e.g.,][]{rogue}, and space weather events (e.g., coronal mass ejections; \cite{Ngwira2013}). These are very difficult to characterize and predict, being exceptionally rare and pathological outliers in the spectrum of weather events. Ensemble forecasting in numerical weather prediction is best suited to estimate statistics of the average or most likely scenarios, and specialized methods are needed to examine the more extreme outlier scenarios.

In this study, we advance an alternative computational approach to predicting and understanding general rare events without sacrificing model fidelity. Our method relies on data generated by a high-fidelity model with a state space with many degrees of freedom $d$, representing, for example, spatial resolution of the primitive equations. In this way, our method is similar to recently introduced reduced order modeling techniques using statistical and machine learning (e.g., \cite{Kashinath2021physics} and references therein). However, in contrast to other data-driven techniques, our approach focuses on directly computing key quantities of interest that characterize the essential predictability of the rare event, rather than trying to capture the full detailed evolution of the system. In particular, we will compute estimators of \emph{statistically optimal forecasts} that are useful for initial conditions somewhere between a ``typical'' configuration $A$ and an ``anomalous'' configuration $B$ that defines the rare event, where typical and anomalous are user-defined. We focus on two forecasts in particular to quantify risk. The \emph{committor} is the probability that a given initial condition evolves directly into $B$ rather than $A$. Given that it does reach $B$ first, the \emph{conditional mean first passage time}, or \emph{lead time}, is the expected time that it takes to get there. The committor appears prominently in the molecular dynamics literature, with some recent applications in geoscience including \cite{Tantet2015,Lucente2019}, and \cite{Finkel2020}, which compute the committor for low-dimensional atmospheric models.

Both quantities depend on the initial condition, defining functions over $d$-dimensional state space that encode important information regarding the fundamental causes and precursors of the rare event. However, ``decoding'' the physical insights is not automatic. With real-time measurement constraints, the risk metrics must be estimated from low-dimensional proxies. Even visualizing them requires projecting down to one or two dimensions. This calls for a principled selection of low-dimensional coordinates which are both physically meaningful and statistically informative for our chosen risk metrics. We address this problem using sparse regression, a simple but easily extensible solution with the potential to inform optimal measurement strategies to estimate risk as precisely as possible under constraints.  

Estimation of the committor and lead time is a challenge. We employ a method that uses a large data set of short-time independent simulations. We represent the committor and lead time as solutions to Feynman-Kac formulae \citep{Oksendal}, which relate long-time forecasts to instantaneous tendencies. These equations are elegant and general, but computationally daunting: in the continuous time and space limit, they become partial differential equations (PDE) with $d$ independent variables---the same as the model state space dimension. It is therefore hopeless to solve the equations using any standard spatial discretization. But, as we demonstrate, the equations can be solved with remarkable accuracy by expanding in a basis of functions informed by the data set. 
 
We illustrate our approach on the highly simplified Holton-Mass model \citep{holton_mass,Christiansen2000} with stochastic velocity perturbations in the spirit of \cite{Birner2008}. The Holton-Mass model is well-understood dynamically in light of decades of analysis and experiments, yet complex enough to present the essential computational difficulties of probabilistic forecasting and test our methods for addressing them. In particular, this system captures the key difficulty in sampling rare events. The vast majority of the time, the system sits in one of two metastable states, characterizing a strong or weak vortex respectively. Extreme events are the infrequent jumps from one state to another. Our computational framework can accurately characterize these rare transitions using only a data set of ``short'' model simulations, short not only compared to the long periods the system sits in one state or the other, but also relative to the timescale of the transition events themselves. In the future, the same methodology could be applied to query the properties of more complex models, such as GCMs, where less theoretical understanding is available. 

In section 2, we review the dynamical model and define the specific rare event of interest. In section 3, we formally define the risk metrics introduced above and visualize the results for the Holton-Mass model, including a discussion of physical and practical insights gleaned from our approach. In section 4 we identify an optimal set of reduced coordinates for estimating risk using sparse regression. These results will provide motivation for the computational method, which we present afterward in section 5 along with accuracy tests. We then lay out future prospects and conclude in section 6. 

\section{Holton-Mass model}
\cite{holton_mass} devised a simple model of the stratosphere aimed at reproducing observed intra-seasonal oscillations of the polar vortex, which they termed ``stratospheric vacillation cycles.'' Earlier SSW models, originating with that of \cite{Matsuno1971}, proposed upward-propagating planetary waves as the major source of disturbance to the vortex. While \cite{Matsuno1971} used impulsive forcing from the troposphere as the source of planetary waves, \cite{holton_mass} suggested that even stationary tropospheric forcing could lead to an oscillatory response, suggesting that the stratosphere can self-sustain its own oscillations. While the Holton-Mass model is meant to represent internal stratospheric dynamics, \cite{Sjoberg2014flux} point out that the stationary boundary condition does not lead to stationary wave activity flux, meaning that even the Holton-Mass model involves some dynamic interaction between the troposphere and stratosphere. Isolating internal from external dynamics is a subtle modeling question, but in the present paper we adhere to the original Holton-Mass framework for simplicity. Our methodology applies equally well to other formulations. 

Radiative cooling through the stratosphere and wave perturbations at the tropopause are the two competing forces that drive the vortex in the Holton-Mass model. %These ``stratospheric vacillation cycles" consist of two phases. First, 
Altitude-dependent cooling relaxes the zonal wind toward a strong vortex in thermal wind balance with a radiative equilibrium temperature field. Gradients in potential vorticity along the vortex, however, can allow the propagation of Rossby waves.  When conditions are just right, a Rossby wave emerges from the tropopause and rapidly propagates upward, sweeping heat poleward and stalling the vortex by depositing a burst of negative momentum. The vortex is destroyed and begins anew the rebuilding process. 

\cite{Yoden1987_bif} found that for a certain range of parameter settings, these two effects balance each other to create two distinct stable regimes: a strong vortex with zonal wind close to the radiative equilibrium profile, and a weak vortex with a possibly oscillatory wind profile. We focus our study on this bistable setting as a prototypical model of atmospheric regime behavior.  The transition from strong to weak vortex state captures the essential dynamics of an SSW. 

The Holton-Mass model takes the linearized quasigeostrophic potential vorticity (QGPV) equation for a perturbation streamfunction $\psi'(x,y,z,t)$ on top of a zonal mean flow $\ov u(y,z,t)$, and projects these two fields onto a single zonal wavenumber $k=2/(a\cos 60^\circ)$ and a single meridional wavenumber $\ell=3/a$, where $a$ is the Earth's radius. This notation is consistent with \cite{holton_mass} and \cite{Christiansen2000}, and we refer the reader to these earlier papers for complete description of the equations and parameters. The resulting ansatz is
\begin{align}
    \ov u(y,z,t)&=U(z,t)\sin(\ell y)\\
    \psi'(x,y,z,t)&=\re\{\Psi(z,t)e^{ikx}\}e^{z/2H}\sin(\ell y)\nonumber
\end{align}
which is fully determined by the reduced state space $U(z,t)$, and $\Psi(z,t)$, the latter being complex-valued. $H$ is a scale height, 7 km. Inserting this into the linearized QGPV equations yields the coupled PDE system
\begin{align}
    &\bigg[-\bigg(\gcal^2(k^2+\ell^2)+\frac14\bigg)+\pdersq{}{z}\bigg]\pder{\Psi}{t}
    \label{eqn:wave}\\
    &\hspace{0.2cm}=\bigg[\bigg(\frac\alpha4-\frac{\alpha_z}{2}-i\gcal^2k\beta\bigg) \nonumber
    -\alpha_z\pder{}{z}-\alpha \pdersq{}{z}\bigg]\Psi \nonumber\\
    &\hspace{0.4cm}+\bigg\{ik\eps\bigg[\bigg(k^2\gcal^2+\frac14\bigg)-\pder{}{z}+\pdersq{}{z}\bigg]U\bigg\}\Psi-ik\eps\pdersq{\Psi}{z}U \nonumber
\end{align}
for $\Psi(z,t)$, and 
\begin{align}
    &\bigg(-\gcal^2\ell^2-\pder{}{z}
    +\pdersq{}{z}\bigg)\pder{U}{t}
    =\big[(\alpha_z-\alpha)U_z^R-\alpha U_{zz}^R\big]
    \label{eqn:meanflow}\\
    &\hspace{0.4cm}-\bigg[(\alpha_z-\alpha)\pder{}{z}+\alpha \pdersq{}{z}\bigg]U+\frac{\eps k\ell^2}2e^z\im\bigg\{\Psi\pdersq{\Psi\str}{z}\bigg\} \nonumber
\end{align}
for $U(z,t)$. Here, $\eps=8/(3\pi)$ is a coefficient for projecting $\sin^2(\ell y)$ onto $\sin(\ell y)$. We have nondimensionalized the equations with the parameter $\gcal^2=H^2N^2/(f_0^2L^2)$, where $N^2=4\times10^{-4}\, \mathrm{s}^{-2}$ is a constant stratification (Brunt-V\"{a}is\"{a}l\"{a} frequency), $f_0$ is the Coriolis parameter, and $L=2.5\times10^5$ m is a horizontal length scale, selected in order to create a homogeneously shaped data set more suited to our analysis. See \cite{holton_mass,Yoden1987_bif,Christiansen2000} for details on parameters. Boundary conditions are prescribed at the bottom of the stratosphere, which in this model corresponds to $z=0$ km, and the top of the stratosphere $z_{top}=70$ km. 
\begin{align}
    \Psi(0,t)=\frac{gh}{f_0}, && \Psi(z_{top},t)=0, \\
    U(0,t)=U^R(0), && \partial_zU(z_{top},t)=\partial_zU^R(z_{top}).\nonumber
\end{align}
The vortex-stabilizing influence is represented by $\alpha(z)$, the altitude-dependent cooling coefficient, and the radiative wind profile $U^R(z)=U^R(0)+\frac{\gamma}{1000}z$ (with $z$ in m), which relaxes the vortex toward radiative equilibrium. Here $\gamma=\bigo(1)$ is the vertical wind shear in m/s/km. The competing force of wave perturbation is encoded through the lower boundary condition $\Psi(0,t)=gh/f_0$. 

\begin{figure*}
    % Snapshots of the two stable states and long run
    \centering
    \includegraphics[trim={0cm 7cm 0cm 0cm},clip,width=0.8\linewidth]{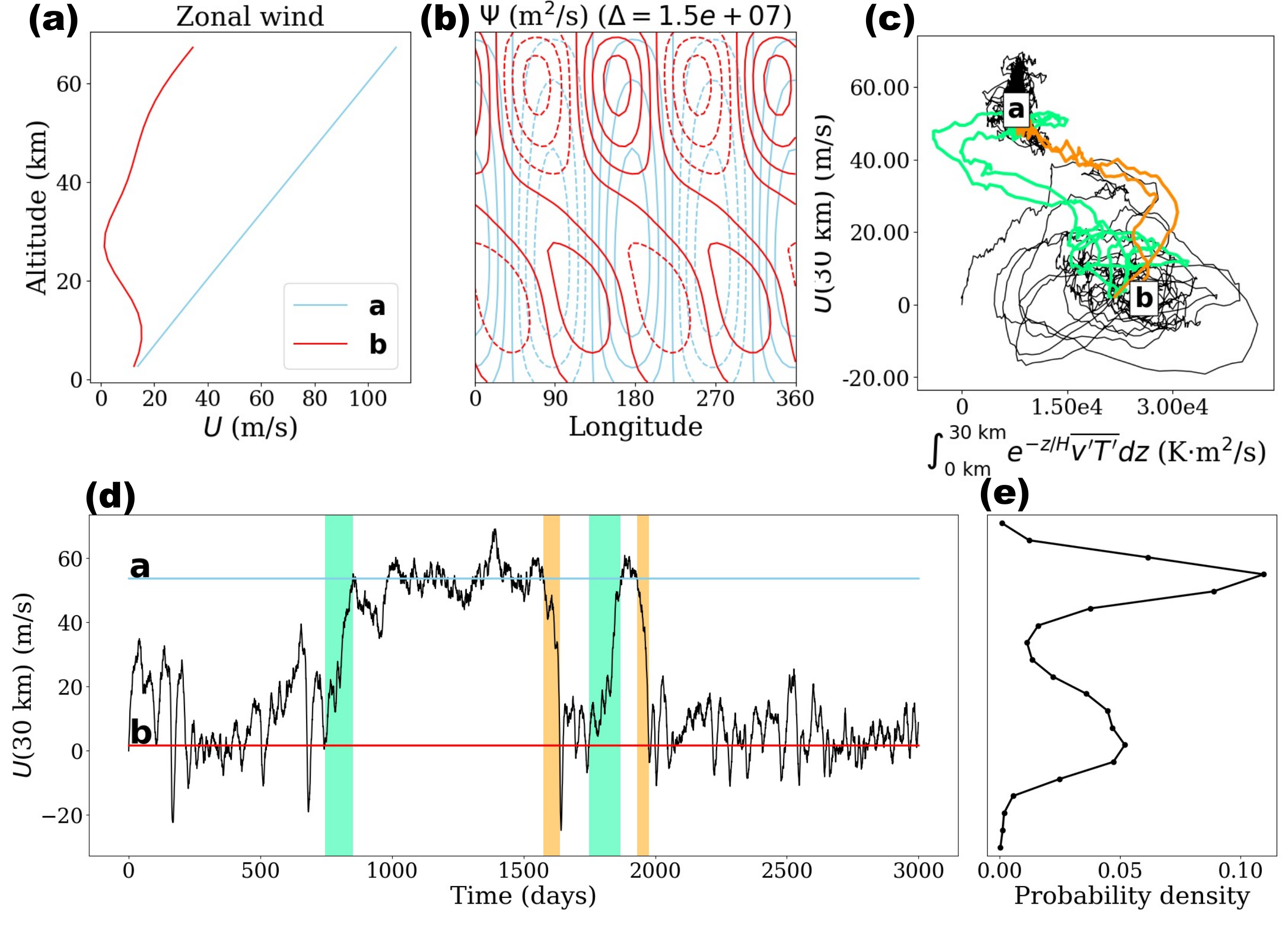}
    \caption{\textbf{Illustration of the two stable states of the Holton-Mass model and transitions between them.} (a) Zonal wind profiles of the radiatively maintained strong vortex (the fixed point $\va$, blue) which increases linearly with altitude, and the weak vortex (the fixed point $\vb$, red) which dips close to zero in the mid-stratosphere. (b) Streamfunction contours are overlaid for the two equilibria $\va$ and $\vb$. (c) Parametric plot of a control simulation in a 2-dimensional state space projection, including two transitions from $A$ to $B$ (orange) and $B$ to $A$ (green). (d) Time series of $U(30$ km) from the same simulation. (e) The steady state density projected onto $U(30\,\mathrm{km})$.}
    \label{fig:equilibria}
\end{figure*}

Detailed bifurcation analysis of the model by both \cite{Yoden1987_bif} and \cite{Christiansen2000} in $(\gamma,h)$ space revealed the bifurcations that lead to bistability, vacillations, and ultimately quasiperiodicity and chaos. Here we will focus on an intermediate parameter setting of $\gamma=1.5$ m/s/km and $h=38.5$ m, where two stable states coexist: a strong vortex with $U$ closely following $U^R$ and an almost barotropic stationary wave, as well as a weak vortex with $U$ dipping close to zero at an intermediate altitude and a stationary wave with strong westward phase tilt. The two stable equilibria, which we call $\va$ and $\vb$, are illustrated in Fig. \ref{fig:equilibria}(a,b) by their $z$-dependent zonal wind and perturbation streamfunction profiles.

The two equilibria can be interpreted as two different winter climatologies, one with a strong vortex and one with a weak vortex susceptible to vacillation cycles. To explore transitions between these two states, we follow \cite{Birner2008} and modify the Holton-Mass equations with small additive noise in the $U$ variable to mimic momentum perturbations by smaller scale Rossby waves, gravity waves, and other unresolved sources. The form of noise will be specified in Eq.~\eqref{eq:noise_form}. 

While the details of the additive noise are ad hoc, the general approach can be more rigorously justified through the Mori-Zwanzig formalism \citep{zwanzig}. Because many hidden degrees of freedom are being projected onto the low-dimensional space of the Holton-Mass model, the dynamics on small observable subspaces can be considered stochastic. This is the perspective taken in stochastic parameterization of turbulence and other high-dimensional chaotic systems \citep{hasselman,Delsole1995,franzke_majda2006,Majda2001stochastic,gottwald2016stochastic}. In general, unobserved deterministic dynamics can make the system non-Markovian, which technically violates the assumptions of our methodology. However, with sufficient separation of timescales the Markovian assumption is not unreasonable. Furthermore, memory terms can be ameliorated by lifting data back to higher-dimensional state space with time-delay embedding \citep{Berry2013timescale,dga,Lin2021koopman}.

We follow \cite{holton_mass} and discretize the equations using a finite-difference method in $z$, with 27 vertical levels (including boundaries). After constraining the boundaries, there are $d=3\times(27-2)=75$ degrees of freedom in the model. \cite{Christiansen2000} investigated higher resolution and found negligible differences. The full discretized state is represented by a long vector
\begin{align}
    \bx(t)=\Big[&\re\{\Psi\}(\Delta z,t),\hdots,\re\{\Psi\}(z_{top}-\Delta z,t),\nonumber\\
    &\im\{\Psi\}(\Delta z,t),\hdots,\im\{\Psi\}(z_{top}-\Delta z,t),\\
    &U(\Delta z,t),\hdots,U(z_{top}-\Delta z,t)\Big]\in\R^d=\R^{75}\nonumber
\end{align}
The deterministic system can be written $d\bx(t)/dt=\bm v(\bx(t))$ for a vector field $\bm v:\R^{d}\to\R^{d}$ specified by discretizing \eqref{eqn:wave} and \eqref{eqn:meanflow}. Under deterministic dynamics, $\bx(t)\to\va$ or $\bx(t)\to\vb$ as $t\to\infty$ depending on initial conditions. The addition of white noise changes the system into an It\^{o} diffusion
\begin{align}
	d\bx(t)=\bm v(\bx(t))\,dt+\bm\sigma(\bx(t))\,d\mathbf{W}(t)
\end{align}
where $\bm\sigma:\R^{d}\to\R^{d\times m}$ imparts a correlation structure to the vector $\mathbf{W}(t)\in\R^{m}$ of independent standard white noise processes. As discussed above, we design $\bm\sigma$ to be a low-rank, constant matrix that adds spatially smooth stirring to only the zonal wind $U$ (not the streamfunction $\Psi$) and which respects boundary conditions at the bottom and top of the stratosphere. Its structure is defined by the following Euler-Maruyama scheme: in a timesetep $\delta t=0.005$ days, after a deterministic forward Euler step we add the stochastic perturbation to zonal wind  on large vertical scales
\begin{align}
    \delta U(z)=\sigma_U\sum_{k=0}^m\eta_k\sin\bigg[\bigg(k+\frac12\bigg)\pi\frac{z}{z_{top}}\bigg]\sqrt{\delta t} \label{eq:noise_form}
\end{align}
where $\eta_k\ (k=0,1,2)$ are independent unit normal samples, $m=2$, and $\sigma_U$ is a scalar that sets the magnitudes of entries in $\bm\sigma$. In terms of physical units,
\begin{align}
    \sigma_U^2=\frac{\ex[(\delta U)^2]}{\delta t}\approx(1\,\text{m/s})^2/\text{day} \label{eq:noise_magnitude}
\end{align}
$\sigma_U$ has units of $(L/T)/T^{1/2}$, where the square-root of time comes from the quadratic variation of the Wiener process. It is best interpreted in terms of the daily root-mean-square velocity perturbation of $1.0$ m/s. We have experimented with this value, and found that reducing the noise level below $0.8$ dramatically reduces the frequency of transitions, while increasing it past $1.5$ washes out metastability. We keep $\sigma_U$ constant going forward as a favorable numerical regime to demonstrate our approach, while acknowledging that the specifics of stochastic parameterization are important in general to obtain accurate forecasts. The resulting matrix $\bm\sigma$ is $75\times3$, with nonzero entries only in the last 25 rows as forcing only applies to $U(z)$. 

A long simulation of the model reveals metastability, with the system tending to remain close to one fixed point for a long time before switching quickly to the other, as shown by the time series of $U(30\,\text{km})$ in panel (d) of Fig. \ref{fig:equilibria}. Panel (e) shows a projection of the steady state distribution, also known as the equilibrium/invariant distribution, of $U$ as a function of $z$. We call this density $\pi(\vx)$, which is a function over the full $d$-dimensional state space. We focus on the zonal wind $U$ at 30 km following \cite{Christiansen2000}, because this is where its strength is minimized in the weak vortex. While the two regimes are clearly associated with the two fixed points, they are better characterized by extended \emph{regions} of state space with strong and weak vortices. We thus define the two metastable subsets of $\R^d$
\begin{align}
    A&=\{\bx:U(\bx)(30\,\mathrm{km})\geq U(\va)(30\, \mathrm{km})=53.8\,\text{m/s}\},\nonumber\\
    B&=\{\bx:U(\bx)(30\,\mathrm{km})\leq U(\vb)(30\, \mathrm{km})=1.75\,\mathrm{m/s}\}.\nonumber
\end{align}
This straightforward definition roughly follows the convention of \cite{cp07}, which defines an SSW as a reversal of zonal winds at 10 hPa. We use 30 km for consistency with Christiansen (2000); this is technically higher than 10 hPa because $z=0$ in the Holton-Mass model represents the tropopause. Our method is equally applicable to any definition, and the results are not qualitatively dependent on this choice. Incidentally, the analysis tools we present may be helpful in distinguishing predictability properties between different definitions. In fact, we will show that the height neighborhood of 20 km is actually more salient for predicting the event than wind at the 30-km level, even when the event is defined by wind at 30 km! This emerges from statistical analysis alone, and gives us confidence that essential SSW properties are stable with respect to reasonable changes in definition.

The orange highlights in Fig. \ref{fig:equilibria} (d) begin when the system exits the $A$ region bound for $B$, and end when the system enters $B$. The green highlights start when the system leaves $B$ bound for $A$, and end when $A$ is reached. Note that $A\to B$ transitions, SSWs, are much shorter in duration than $B\to A$ transitions. Fig. \ref{fig:equilibria} (c) shows the same paths, but viewed parametrically in a two-dimensional state space consisting of integrated heat flux or IHF $\int_{0\ \mathrm{km}}^{30\,\mathrm{km}}e^{-z/H}\ov{v'T'}\,dz$, and zonal wind $U$(30 km). IHF is an informative number because it captures both magnitude and phase information of the streamfunction in the Holton-Mass model:
\begin{align}
    \mathrm{IHF}=\int_{0\ \mathrm{km}}^{30\ \mathrm{km}}e^{-z/H}\ov{v'T'}\,dz\propto\int_{0\ \mathrm{km}}^{30\ \mathrm{km}}|\Psi|^2\pder{\varphi}{z}\,dz
\end{align}
where $\varphi$ is the phase of $\Psi$. 
The $A\to B$ and $B\to A$ transitions are again highlighted in orange and green respectively, showing geometrical differences between the two directions. We will refer to the $A\to B$ transition as an SSW event, even though it is more accurately a transition between climatologies according to the Holton-Mass interpretation. The $B\to A$ transition is a vortex restoration event. Our focus in this paper is on predicting these transition events (mainly the $A\to B$ direction) and monitoring their progress in a principled way. In the next section we explain the formalism for doing so.

\section{Forecast functions: the committor and lead time statistics}

\subsection{Defining risk and lead time}
We will introduce the quantities of interest by way of example. First, suppose the stratosphere is observed in an initial state $\bx(0)=\vx$ that is neither in $A$ nor $B$, so $U(\vb)(30\,\mathrm{km})<U(\vx)(30 \,\mathrm{km})<U(\va)(30\,\mathrm{km})$ and the vortex is somewhat weakened, but not completely broken down. We call this intermediate zone $D=(A\cup B)\cm$ (the complement of the two metastable sets). Because $A$ and $B$ are attractive, the system will soon find its way to one or the other at the \emph{first-exit time} from $D$, denoted
\begin{align}
    \tau_{D\cm}=\min\{t\geq0: \bx(t)\in D\cm\}
    \label{eqn:stopping_time}
\end{align}
Here, $D\cm$ emphasizes that the process has left $D$, i.e., gone to $A$ or $B$. The first-exit location $\bx(\tau_{D\cm})$ is itself a random variable which importantly determines how the system exits $D$: either $\bx(\tau_{D\cm})\in A$, meaning the vortex restores to radiative equilibrium, or $\bx(\tau_{D\cm})\in B$, meaning the vortex breaks down into vacillation cycles. A fundamental goal of forecasting is to determine the probabilities of these two events, which naturally leads to the definition of the (forward) committor function
\begin{align}
    \qp(\vx)&=
    \begin{cases}
        \pr_\vx\{\bx(\tau_{D\cm})\in B\} & \vx\in D=(A\cup B)\cm\\
        0 & \vx\in A\\
        1 & \vx\in B
    \end{cases}
    \label{eqn:committor_definition}
\end{align}
where the subscript $\vx$ indicates that the probability is conditional on a fixed initial condition $\bx(0)=\vx$, i.e., $\pr_\vx\{\cdot\}=\pr\{\cdot|\bx(0)=\vx\}$. The superscript ``+'' distinguishes the forward committor from the \emph{backward committor}, an analogous quantity for the time-reversed process which we do not use in this paper. Throughout, we will use capital $\bx(t)$ to denote a stochastic process, and lower-case $\vx$ to represent a specific point in state space, typically an initial condition, i.e., $\bx(0)=\vx$. Both are $d=75$-dimensional vectors. 

The committor is the probability that the system will be in state $B$ (the disturbed state) next rather than $A$ (the strong vortex state). Hence $\qp(\vx)=0$ if you start in $A$, and is 1 if you are already in $B$. In between (i.e., when $\vx\in D$), $\qp(\vx)$ tells you the probability that you will first go to $B$ rather than to $A$. That is, $\qp(\vx)$ tells you the probability that an SSW will happen.

Another important forecasting quantity is the lead time to the event of interest. While the forward committor reveals the probability of experiencing vortex breakdown \emph{before} returning to a strong vortex, it does not say how long either event will take. Furthermore, even if the vortex is restored first, how long will it be until the next SSW does occur? The time until the next SSW event is denoted $\tau_B$, again a random variable, whose distribution depends on the initial condition $\vx$. We call $\ex_\vx[\tau_B]$ the \emph{mean first passage time} (MFPT) to $B$. Conversely, we may ask how long a vortex disturbance will persist before normal conditions return; the answer (on average) is $\ex_\vx[\tau_A]$, the mean first passage time to $A$. 
These same quantities have been calculated previously in other simplified models, e.g. \cite{Birner2008} and \cite{Esler2019}. 

$\ex_\vx[\tau_B]$ has an obvious shortcoming: it is an average over all paths starting from $\vx$, including those which go straight into $B$ (i.e., an orange trajectory in Fig. \ref{fig:equilibria}c,d) and the rest which return to $A$ i.e., a green trajectory) and linger there, potentially for a very long time, before eventually re-crossing back into $B$. It is more relevant for near-term forecasting to condition $\tau_B$ on the event that an SSW is coming before the strong vortex returns. For this purpose, we introduce the \emph{conditional} mean first passage time, or lead time, to $B$:
\begin{align}
    \lt(\vx):=\ex_\vx[\tau_B|\tau_B<\tau_A]
\end{align} 
which quantifies the suddenness of SSW. 

All of these quantities can, in principle, be estimated by direct numerical simulation (DNS). For example, suppose we observe an initial condition $\bx(0)=\vx$ in an operational forecasting setting, and wish to estimate the probability and lead time for the event of next hitting $B$. We would initialize an ensemble $\{\bx_n(0)=\vx,n=1,\hdots,N\}$ and evolve each member forward in time until it hits $A$ or $B$ at the random time $\tau_n$. In an explicitly stochastic model, random forcing would drive each member to a different fate, while in a deterministic model their initial conditions would be perturbed slightly. To estimate the committor to $B$, we could calculate the fraction of members that hit $B$ first. Averaging the arrival times ($\tau_n$), over only those members gives an estimate of the lead time to $B$. For a single initial condition $\vx$ reasonably close to $B$, DNS may be the most economical. But how do we systematically compute $\qp(\vx)$ over all of state space (here 75 variables, but potentially billions of variables in a GCM or other state-of-the-art forecast system)? 

For this more ambitious goal, DNS is prohibitively expensive.  By definition, transitions between $A$ and $B$ are infrequent.  Therefore, if starting from $\vx$ far from $B$, a huge number of sampled trajectories ($N$) will be required to observe even a small number ending in $B$, and they may take a long time to get there. If instead we could precompute these functions offline over all of state space, the online forecasting problem would reduce to ``reading off'' the committor and lead time with every new observation. Achieving this goal is the key point of our paper, and we achieve this using the dynamical Galerkin approximation, or DGA, recipe described by \cite{dga}.

A brute force way to estimate these functions is to integrate the model for a long time until it reaches statistical steady state, meaning it has explored its attractor thoroughly according to the steady state distribution. After long enough, it will have wandered close to every point $\vx$ sufficiently often to estimate $\qp(\vx)$ and $\lt$ robustly as in DNS. We have performed such a ``control simulation'' of $5\times10^5$ days for validation purposes, but our main contribution in this paper is to compute the forecast functions using only \emph{short} trajectories with DGA, allowing for massive parallelization. However, we will defer the methodological details to Section 5, and first justify the effort with some results. We visualize the committor and lead time computed from short trajectories and elaborate on their interpretation, utility, and relationship to ensemble forecasting methods. 

\subsection{Steady state distribution}
Before visualizing the committor and lead time, it will be helpful to have a precise notion of the steady state distribution, denoted $\pi(\vx)$, a probability density that describes the long-term behavior of a stochastic process $\bx(t)$. Assuming the system is ergodic, averages over time are equivalent to averages over state space with respect to $\pi$. That is, for any well-behaved function $g:\R^d\to\R$,
\begin{align}
    \ang{g}_\pi:=\lim_{T\to\infty}\frac1{T}\int_0^Tg(\bx(t))\,dt=\int_{\R^d}g(\vx)\pi(\vx)\,d\vx \label{eq:ergodicity_definition}
\end{align}
For example, if $g(\vx)=\ind_S(\vx)$ (an indicator function, which is 1 for $\vx\in S\subset\R^d$ and 0 for $\vx\notin S$), Eq.~\eqref{eq:ergodicity_definition} says that the fraction of time spent in $S$ can be found by integrating the density over $S$. The density peaks in Fig. \ref{fig:equilibria}(d) indicates clearly that the neighborhoods of $\va$ and $\vb$ are two such regions with especially large probability under $\pi$. Note that both sides of~\eqref{eq:ergodicity_definition} are independent of the initial condition, which is forgotten eventually. Short-term forecasts are by definition out-of-equilibrium processes, depending critically on initial conditions; however, $\pi(\vx)$ is important to us here as a ``default'' distribution for missing information. If the initial condition is only partially observed, e.g. in only one coordinate, we have no information about the other $d-1$ dimensions, and in many cases the most principled tactic is to assume those other dimensions are distributed according to $\pi$, conditional on the observation.

\subsection{Visualizing committor and lead times}

The forecasts $\qp(\vx)$ and $\lt(\vx)$ are functions of a high-dimensional space $\R^d$. However, these degrees of freedom may not all be ``observable'' in a practical sense, given the sparsity and resolution limits of weather sensors, and visualizing them requires projecting onto reduced-coordinate spaces of dimension 1 or 2. We call these ``collective variables'' (CVs) following chemistry literature \citep[e.g.,][]{Noe2017cv}, and denote them as vector-valued functions from the full state space to a reduced space, $\bm\theta:\R^d\to\R^k$, where $k=1$ or $2$. For instance, Fig. \ref{fig:equilibria} (c) plots trajectories in the CV space consisting of integrated heat flux and zonal wind at 30 km: $\bm\theta(\vx)=\Big(\int_{0\ \mathrm{km}}^{30\,\mathrm{km}}e^{-z/H}\ov{v'T'}\,dz,U(30\,\mathrm{km})\Big)$. The first component is a nonlinear function involving products of $\re\{\Psi\}$ and $\im\{\Psi\}$, while the second component is a linear function involving only $U$ at a certain altitude. For visualization in general, we have to approximate a function $F:\R^d\to\R$, such as the committor or lead time, as a function of reduced coordinates. That is, we wish to find $f:\R^k\to\R$ such that $F(\vx)\approx f(\bm\theta(\vx))$. Given a fixed CV space $\bm\theta$, an ``optimal'' $f$ is chosen by minimizing some function-space metric between $f\circ\bm\theta$ and $F$. 

\begin{figure*}
    % 1-dimensional projections
    \centering
    \includegraphics[trim={0cm 0cm 0cm 0cm},clip,width=0.9\linewidth]{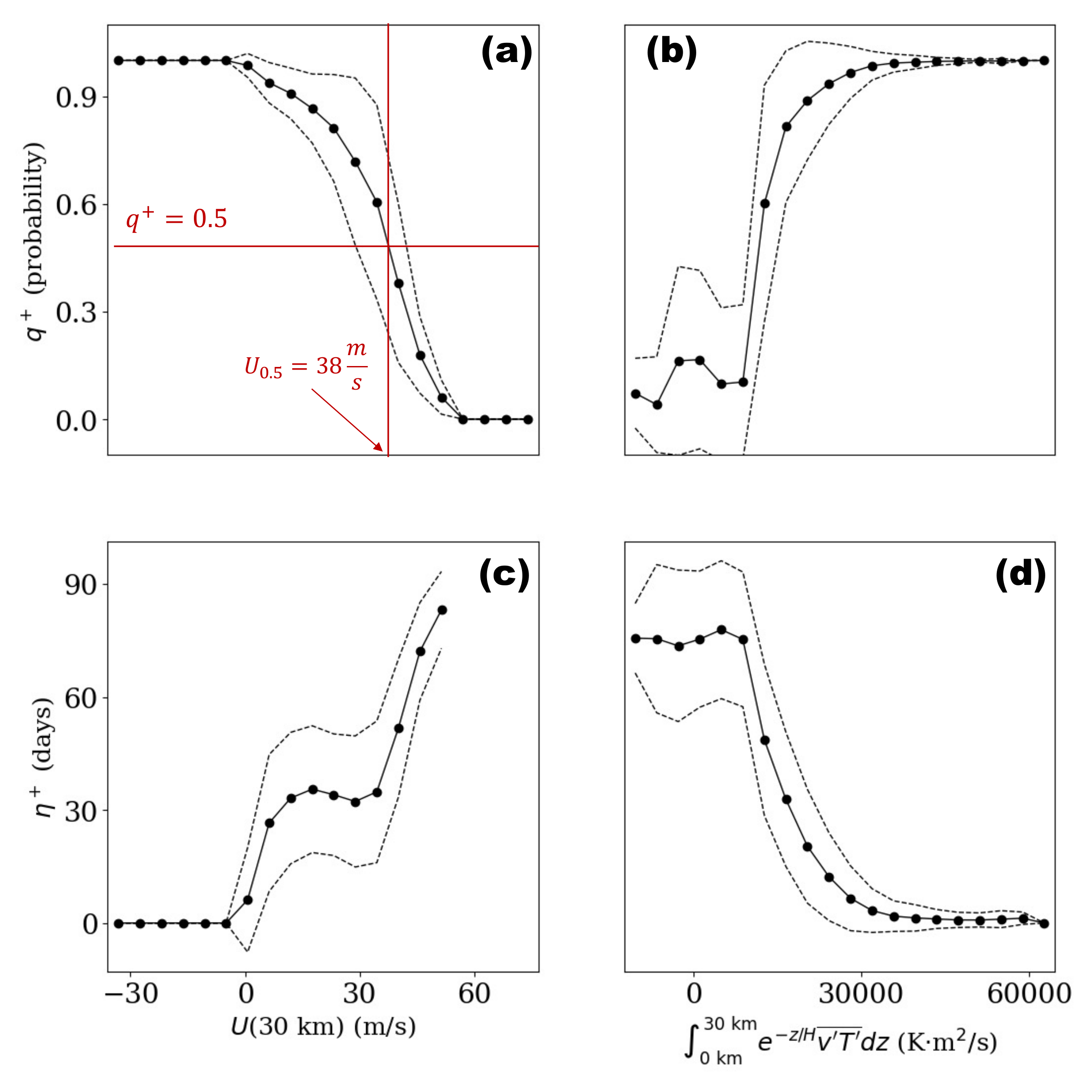}
    \caption{\textbf{One-dimensional projections of the forward committor (first row) and lead time to $B$ (second row)}. These functions depend on all $d=75$ degrees of freedom in the model, but we have averaged across $d-1=74$ dimensions to visualize them as rough functions of two single degrees of freedom: $U(30$ km) (first column) and integrated heat flux up to 30 km, IHF (second column). Panel (a) additionally marks the $\qp=\frac12$ threshold and the corresponding value of zonal wind.} 
    \label{fig:committor_mfpt_1d}
\end{figure*}

A natural choice is the mean-squared error weighted by the steady state distribution $\pi$, so the projection problem is to minimize over functions $f:\R^k\to\R$ the penalty
\begin{align}
    S[f;\bm\theta]:&=\norm{f\circ\bm\theta-F}_{L^2(\pi)}^2 \nonumber\\
    &=\int_{\R^d}\Big[f(\bm\theta(\vx))-F(\vx)\Big]^2\pi(\vx)\,d\vx. \label{eq:projection_formula_0}
\end{align}
The optimal $f$ for this purpose is the conditional expectation
\begin{align}
    f(\vy)&=\ex_{\bx\sim\pi}[F(\bx)|\bm\theta(\bx)=\vy]\nonumber\\
    &=\lim_{|d\vy|\to0}\frac{\int f(\vx)\ind_{d\vy}(\bm\theta(\vx))\pi(\vx)\,d\vx}{\int\ind_{d\vy}(\bm\theta(\vx))\pi(\vx)\,d\vx}    
    \label{eq:projection_formula_1}
\end{align}
where $d\vy$ is a small neighborhood about $\vy$ in CV space $R^k$. The subscript $\bx\sim\pi$ means that the expectation is with respect to a random variable $\bx$ distributed according to $\pi(\vx)$, i.e., at steady state. Fig. \ref{fig:committor_mfpt_1d} uses this formula to display one-dimensional projections of the committor (first row) and lead time (second row), as well as the one-standard deviation envelope incurred by projecting out the other 74 degrees of freedoom. This ``projection error'' is defined as the square root of the conditional variance 
\begin{align}
   V_F(\vy)=\ex_{\bx\sim\pi}\Big[\Big(F(\bx)-f(\vy)\Big)^2\Big|\bm\theta(\bx)=\vy\Big].
\end{align}
Each quantity is projected onto two different one-dimensional CVs: $U(30\ \mathrm{km})$ (first column) and IHF (second column). In panel (a), for example, we see the committor is a decreasing function of $U$: the weaker the wind, the more likely a vortex breakdown. Moreover, the curve provides a conversion factor between risk (as measured by probability) and a physical variable, zonal wind. An observation of $U(30\ \mathrm{km})=38$ m/s implies a 50\% chance of vortex breakdown. The variation in slope also tells us that a wind reduction from 40 m/s to 30 m/s represents a far greater increase in risk than a reduction from 30 m/s to 20 m/s. Meanwhile, panel (b) shows the committor to be an increasing function of IHF, since SSW is associated with large wave amplitude and phase lag. However, IHF seems inferior to zonal wind as a committor proxy, as a small change in IHF from $\sim15000$ to $\sim20000$ K$\cdot$m$^2$/s corresponds to a sharp increase in committor from nearly zero to nearly one. In other words, knowing only IHF doesn't provide much useful information about the threat of SSW until it is already virtually certain. The dotted envelope is also wider in panel (b) than (a), indicating that projecting the committor onto IHF removes more information than projecting onto $U$. While the underlying noise makes it impossible to divine the outcome with certainty from \emph{any} observation, the projection error clearly privileges some observables over others for their predictive power.

In panels (c) and (d), the lead time is seen to have the opposite overall trend as the committor: the weaker the wind, or the greater the heat flux, the closer you are on average to a vortex breakdown. $\lt(\vx)$ is  not defined when wind is strongest, as $\vx\in A$ and so $\qp(\vx)=0$. However, an interesting exception to the trend occurs in the range $10\ \mathrm{m/s}\leq U\leq40\ \mathrm{m/s}$: the expected lead time stays constant or slightly \emph{decreases} as zonal wind increases, and the projection error remains large. This means that while the probability of vortex breakdown increases rapidly from 50\% to 90\%, the time until vortex breakdown remains highly uncertain. To resolve this seeming paradox, we will have to visualize the joint variation of $\qp$ and $\lt$.

\begin{figure*}
    % Committor demonstration
    \centering
    \includegraphics[trim={0 0cm 0 0},clip,width=0.99\linewidth]{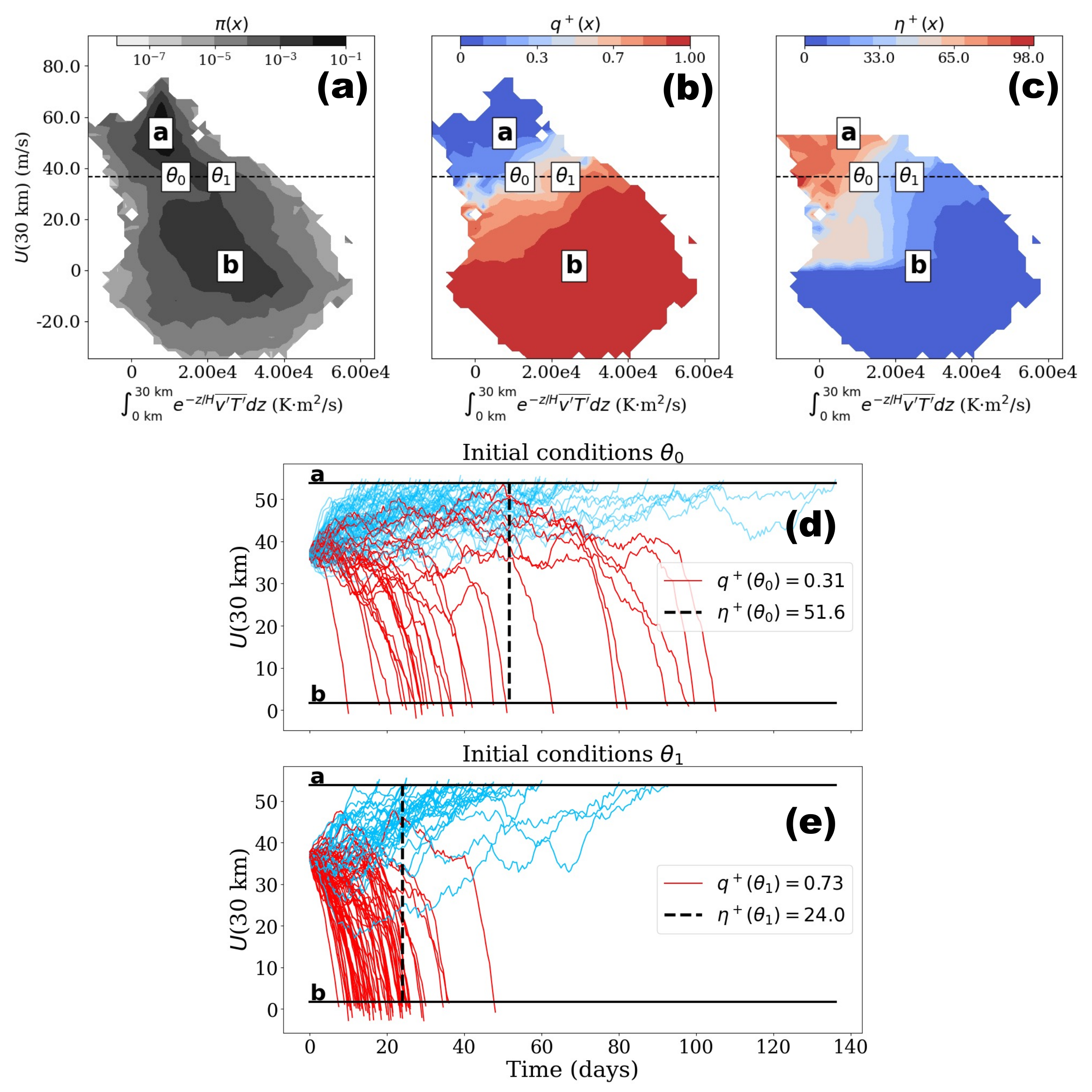}
    \caption{\textbf{The density, committor, and lead time as functions of zonal wind and integrated heat flux.} Panel (a) projects the steady state distribution $\pi(\vx)$ onto the two-dimensional subspace ($U$,IHF) at 30 km. The white regions surrounding the gray are unphysical states with negligible probability. Panels (b) and (c) display the committor and lead time in the same space. A horizontal transect marks the level $U$(30 km) = 38.5 m/s, where $\qp$ according to $U$ only is 0.5. Panels (d) and (e) show ensembles initialized from two points $\bm\theta_0$ and $\bm\theta_1$ along the transect, verifying that their committor and lead time values differ from their values according to $U$, in a way that is predictable due to considering IHF in addition to $U$.}
    \label{fig:transition_ensemble}
\end{figure*}

It is of course better to consider multiple observables at once. Fig. \ref{fig:transition_ensemble} shows the information gained beyond observing $U(30\ \mathrm{km})$ by incorporating IHF as a second observable. In the top row we project $\pi$, $\qp$, and $\lt$ onto the two-dimensional subspace, revealing structure hidden from view in the one-dimensional projections. Panel (a) is a 2-dimensional extension of Fig. \ref{fig:equilibria}(d), with density peaks visible in the neighborhoods of $\va$ and $\vb$. The white space surrounding the gray represents physically insignificant regions of state space that was not sampled by the long simulation. The same convention holds for the following two-dimensional figures. The committor is displayed in panel (b) over the same space. It changes from blue at the top (an SSW is unlikely) to red at the bottom (an SSW is likely), bearing out the negative association between $U$ and $\qp$. However, there are non-negligible horizontal gradients that show that IHF plays a role, too. Likewise, the lead time in panel (c) decreases from $\sim90$ days near $\va$ to 0 days near $\vb$, when the transition is complete. Here, IHF appears even more critically important for forecasting how the event plays out, as gradients in $\eta^+$ are often completely horizontal.

A horizontal dotted line in Fig. \ref{fig:transition_ensemble}(a-c) marks the 50\% risk level $U(30\ \mathrm{km})=38$ m/s, but the committor varies along it from low risk at the left to high risk at the right: we show this concretely by selecting two points $\bm{\theta}_0$ and $\bm{\theta}_1$ along the line. According to $U$ alone, i.e., the curve in Fig. \ref{fig:committor_mfpt_1d}(a), both would have the same committor of 0.5. According to both $U$ and IHF together, i.e., the two-dimensional heat map in Fig. \ref{fig:transition_ensemble}(b), they have very different probabilities of $\qp(\bm\theta_0)=0.31$ and $\qp(\bm\theta_1)=0.73$: an SSW is more than twice as likely to occur from starting point $\bm\theta_1$ as $\bm\theta_0$. 

While those committor values come from the DGA method to be described in Section 5, we confirm them empirically by plotting an ensemble of 100 trajectories originating from each of the two initial conditions in panels (d) and (e) below, coloring $A$-bound trajectories blue and $B$-bound trajectories red. Only 28\% of the sampled trajectories through $\bm\theta_0$ exhibit an SSW, next going to state $B$, while 68\% of the integrations from $\bm\theta_1$ end at $B$. In both cases, the heatmaps and ensemble sample means roughly match. The small differences between the projected committor and the empirical ``success'' rate of trajectories arises both from errors in the DGA calculation (which we analyze in section 5) and the finite size of the ensemble. 

The lead time prediction is improved similarly by incorporating the second observable. According to $U$ alone, Fig. \ref{fig:committor_mfpt_1d} predicts a lead time of $40$ days for both $\bm\theta_0$ and $\bm\theta_1$. Considering IHF additionally, the two-dimensional heat map in Fig. \ref{fig:transition_ensemble} predicts a lead time of 52 days and 24 days for $\bm\theta_0$ and $\bm\theta_1$, respectively. Referring to the ensemble from $\bm\theta_1$ in panels (d) and (e), the arrival times of red trajectories to $B$ provide a discrete sampling of the lead time distributions of $\tau_B|\tau_B<\tau_A$. The sample means are 50 and 32 days respectively from $\bm\theta_0$ and $\bm\theta_1$, again roughly matching with our predictions. 

These two-dimensional projections still leave out 73 remaining dimensions, which we could incorporate to make the forecasts even better. After accounting for all 75 dimensions, we would obtain the full committor function $\qp:\R^d\to\R$. This is still a probability, i.e., an expectation over the unresolved turbulent processes and uncertain initial condition. Low-dimensional committor projections simply treat the projected-out dimensions as random variables sampled according to $\pi$. Whether projected to a space of 1 or 75 dimensions, the committor is the function of that space that is closest, in the mean-square sense, to the binary indicator $\ind_B(\bx(\tau))$; this is the defining characteristic of conditional expectation \citep{Durrett}. In the case that the system does hit $B$ next, the lead time is closest in the mean-square sense to $\tau_B$. 

While high-dimensional systems offer many coordinates to choose from, we argue that the committor and lead time are the most important nonlinear coordinates to monitor for forecasting purposes. We will explore their relationship in the next subsection. Although both encode some version of proximity to SSW, they are independent variables which deserve separate consideration.

% ----------------------------------------------

\subsection{Relationship between risk and lead time}
A forecast is most useful if it comes sufficiently early (to leave some buffer time before impact) and is sufficiently precise to time your response. For example, in June we can say with certainty it will snow next winter in Minnesota. To be useful, we want to know the date of the first snow as early as possible. By relating levels of risk (quantified by $\qp$) and lead time (quantified by $\lt$), we can now assess the limits of early prediction. Such a relationship would answer two questions: for an SSW transition, (1) how far in advance will we be aware of it with some prescribed confidence, say 80\%? (2) given some prescribed lead time, say 42 days, how aware or ignorant could we be of it?

\begin{figure*}
    % Inversion: zonal wind and IHF as functions of committor and conditional passage time
    \centering
    \includegraphics[trim={0cm 0cm 0cm 0cm},clip,width=0.8\linewidth]{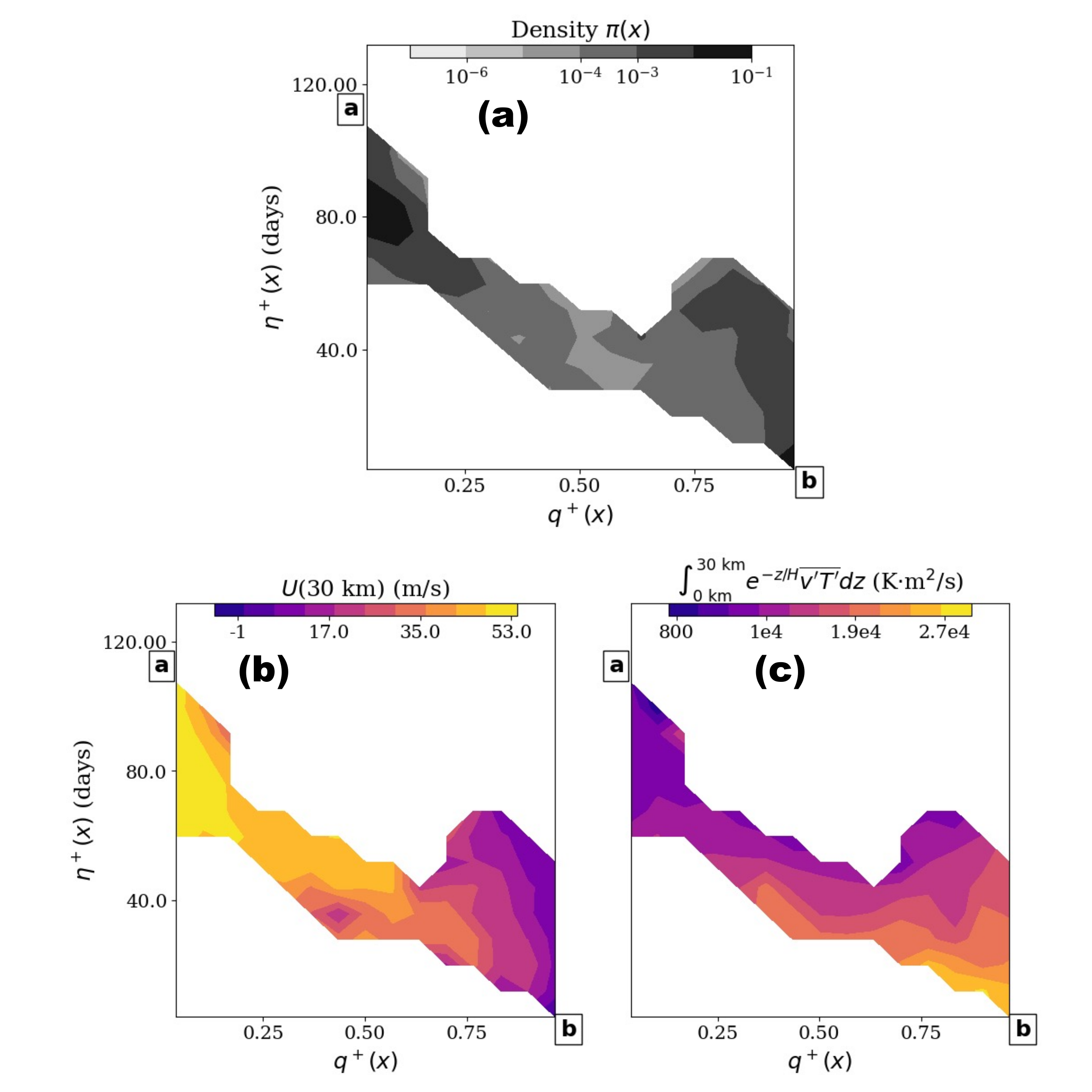}
    \caption{\textbf{Committor and lead time as independent coordinates.} This figure inverts the functions in Fig. \ref{fig:transition_ensemble}, considering the zonal wind and integrated heat flux as functions of committor and lead time. The two-dimensional space they span is the essential goal of forecasting. Panel (a) shows the steady state distribution on this subspace, which is peaked near $\va$ and $\vb$ (darker shading), weaker in the "bridge" region between them, and completely negligible the white regions unexplored by data. Panels (b) and (c) display zonal wind and heat flux in color as functions of the committor and lead time.}
    \label{fig:committor_mfpt_scatter}
\end{figure*}

The committor and lead time have an overall negative relationship, but they do not completely determine each other, as the contours in Fig. \ref{fig:transition_ensemble}(a,b) do not perfectly line up. We treat them as independent variables in Fig. \ref{fig:committor_mfpt_scatter}, which maps zonal wind and IHF as functions of the coordinates $\qp$ and $\lt$ in an inversion of Fig. \ref{fig:transition_ensemble}. The density $\pi(\vx)$ projected on this space in 4(a) shows again a bimodal structure around $\va$ and $\vb$, which occupy opposite corners of this space by construction. Meanwhile, zonal wind and IHF are indicated by the shading in panels (b) and (c). The bridge between $\va$ and $\vb$ is not a narrow band, but rather includes a curious high-committor, high-lead time branch which seems paradoxical: points at $\qp=0.9$ have a greater spread in $\lt$ than points at $\qp=0.5$, contrary to the intuition that closeness to $B$ in probability means closeness in time. The color shading shows that $\qp$ is strongly associated with $U$(30 km), while $\lt$ is more strongly associated with IHF(30 km). In particular the horizontal contours in panel (c) show that the large spread in lead time near $B$ is due almost completely to variation in IHF. In other words, the system can be highly committed to $B$ with a low zonal wind, but if IHF is low, it may take a long time to get there. We can also see this from the lower-left region of Fig. \ref{fig:transition_ensemble}(a) and (b), where committor is high and lead time is high. 

There are two complementary explanations for this phenomenon. First, the low-$U$, low-IHF region of state space corresponds to a temporary restoration phase in a vacillation cycle, which delays the inevitable collapse of zonal wind below the threshold defining $B$. In fact, the ensemble of pathways starting from $\bm\theta_0$ in Fig. \ref{fig:transition_ensemble}(c) has several members whose zonal wind either stagnates at medium strength, or dips low and partially restores before finally plunging all the way down. The second explanation is that many of these partial restoration events are not part of an $A\to B$ transition, but rather a $B\to B$ transition. In a highly irreversible system such as the Holton-Mass model, these two situations are quite dynamically distinct. To distinguish them using DGA, we would have to account for the \emph{past} as well as the future, calculating backward-in-time forecasts such as the backward committor $\qm(\vx)=\pr_\vx\{\bx(\tau^-)\in A\}$, where $\tau^-<0$ is the most-recent hitting time. Backward forecasts will be analyzed thoroughly in a forthcoming paper, but they are beyond the scope of the present one.

In summary, $\qp$ and $\lt$ are principled metrics to inform preparation for extreme weather. For example, a threatened community might decide in advance to start taking action when an event is very likely, $\qp\geq0.8$, and somewhat imminent, $\lt\leq10$ days, or rather, when an event is somewhat likely, $\qp\geq0.5$, and very imminent, $\lt\leq3$ days. Because of partial restoration events, the committor does not determine the lead time or vice versa, and so a good real-time disaster response strategy should take both of them into account, defining an ``alarm threshold'' that is not a single number, but some function of both the committor and lead time. This idea is similar in spirit to that of the Torino scale, which assigns a single risk metric to an asteroid or comet impacts based on both probability and severity \citep{Binzel2000torino}.  Of course, after many near-SSW events, a lot of material damage may have already occurred, which may be a reason to define a higher threshold for the definition of $B$, or even a continuum for different severity levels of SSW. We emphasize that the choice of $A$, $B$ and alarm thresholds are more of a community and policy decision than a scientific one. The strength of our approach is that it provides a flexible numerical framework to quantify and optimize the consequences of those decisions. 

\section{Sparse representation of the committor}

The committor projections showed give only an impression of its high-dimensional structure. While Eq.~\eqref{eq:projection_formula_1} says how to optimally represent the committor over a given CV subspace, optimizing $S[f;\bm\theta]$ over $f$, it does not say which subspace $\bm\theta$ is optimal. If the committor does admit a sparse representation, we could specifically target observations on these high-impact signals. In this section we address this much harder problem of optimizing $S[f;\bm\theta]$ over subspaces $\bm\theta$. 

The set of CV spaces is infinite, as observables $\bm\theta$ can be arbitrarily complex nonlinear functions of the basic state variables $\vx$. Machine learning algorithms such as artificial neural networks are designed exactly for that purpose: to represent functions nonparametrically from observed input-output pairs. However, to keep the representation interpretable, we will restrict ourselves to physics-informed input features based on the Eliassen-Palm (EP) relation, which relates wave activity, PV fluxes and gradients, and heating source terms in a conservation equation. From \cite{Yoden1987_dyn}, the EP relation for the Holton-Mass model takes the form
\begin{align}
    \partial_t\bigg(\frac{q'^2}2\bigg)&+(\partial_y\ov q)
    %\bmt
    \rho_s\inv\nabla\cdot\bm F\nonumber\\
    %\ov{v'q'}\\
    %\frac{R}{Hf_0}\rho_s\inv\ov{v'T'}
    %\emt
    &=-\fn\rho_s\inv\ov{q'\partial_z(\alpha\rho_s\partial_z\psi')} 
    \label{eqn:ep_relation}\\
    \text{ where }\bm F=(-\rho_s\ov{u'v'})&\jhat+(\rho_s\ov{v'\partial_z\psi'})\khat\nonumber
\end{align}
The EP flux divergence has two alternative expressions: $\rho_s\inv\nabla\cdot\bm F=\ov{v'q'}=\rho_s\inv\frac{R}{Hf_0}\partial_z\big[\rho_s\ov{v'T'}\big]$. If there were no dissipation $(\alpha=0)$ and the background zonal state were time-independent ($\partial_t\ov q=0$), dividing both sides by $\partial_y\ov q$ would express local conservation of wave activity $\acal=\rho_s\ov{q'^2}/(2\partial_y\ov q)$. Neither of these is exact in the stochastic Holton-Mass model, so we use the quantities in Eq.~\eqref{eqn:ep_relation} as diagnostics: enstrophy $\ov{q'^2}$, PV gradient $\partial_y\ov q$, PV flux $\ov{v'q'}$, and heat flux $\ov{v'T'}$. Each field is a function of $(y,z)$ and takes on very different profiles for the states  $\va$ and $\vb$, as found by \cite{Yoden1987_dyn}. A transition from $A$ to $B$, where the vortex weakens dramatically, must entail a reduction in $\partial_y\ov q$ and a burst in positive $\ov{v'T'}$ (negative $\ov{v'q'}$) as a Rossby wave propagates from the tropopause vertically up through the stratosphere and breaks. This is the general physical narrative of a sudden warming event, and these same fields might be expected to be useful observables to track for qualitative understanding and prediction. For visualization, we have found $U(30\,km)$ and IHF(30 km) $=\int_{0\,\mathrm{km}}^{30\,\mathrm{km}}e^{-z/H}\ov{v'T'}\,dz$ to be particularly helpful. However, this doesn't necessarily imply they are optimal predictors of $\qp$, and regression is a more principled way to find them.

\begin{figure*}
    % z family
    \centering
    \includegraphics[trim={6 0cm 6cm 0},clip,width=0.8\linewidth,]{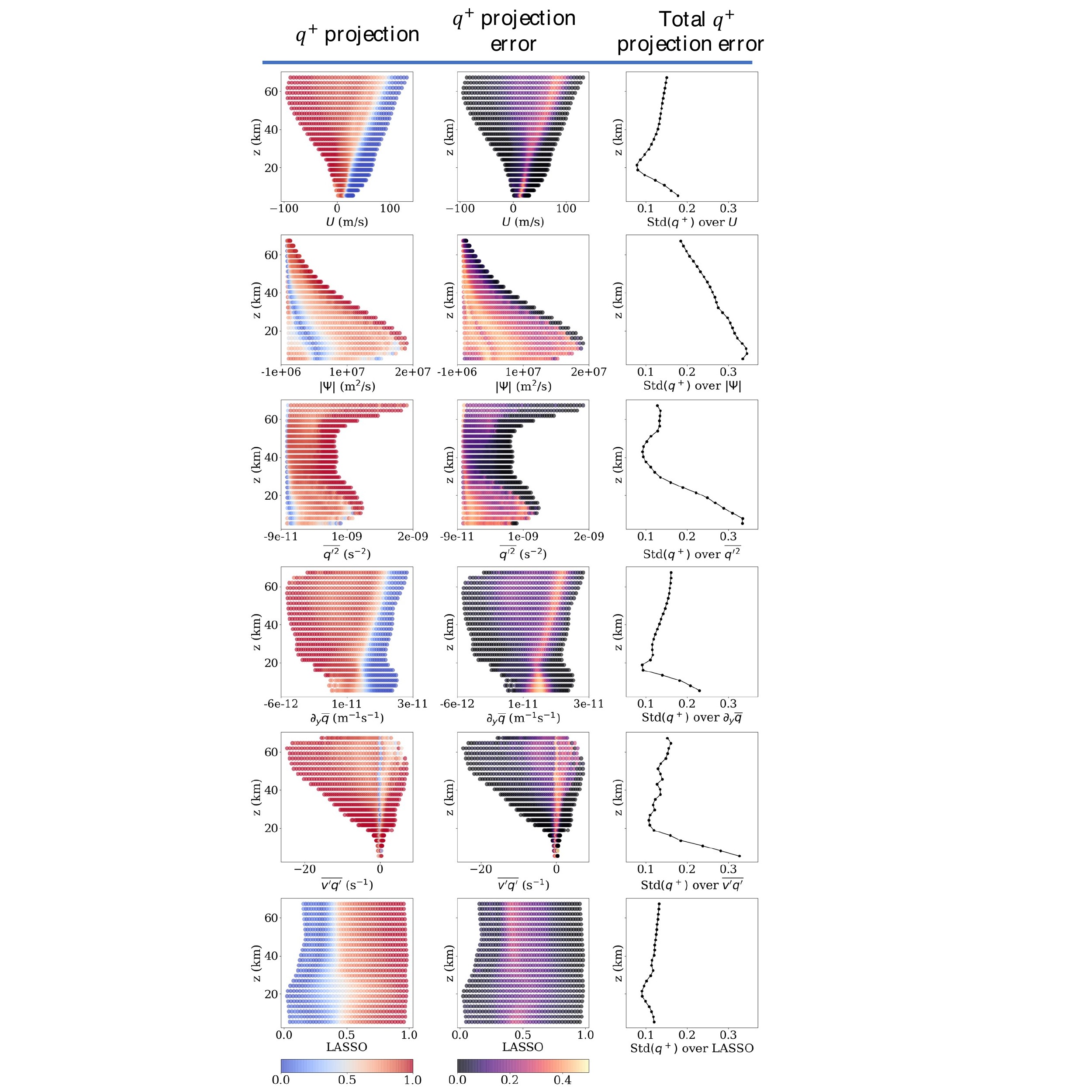}
    \caption{\textbf{Projection of the forward committor onto a large collection of altitude-dependent physical variables}. The top left panel shows heatmaps of $\qp$ as a function of $U$ and $z$; white regions denote where $U(z)$ is negligibly observed. The top middle panel shows the standard deviation in $\qp$ as a function of $U$ and $z$; this uncertainty stems from the remaining 74 model dimensions. The right-hand panel displays the total mean-squared error due to the projection for each altitude, i.e., $\sqrt{S[f;\bm\theta]}$ from Eq.~\eqref{eq:projection_formula_0}. A low value indicates that this level is ideal for prediction. The following rows show the same quantities for other physical variables: streamfunction magnitude, eddy enstrophy, background PV gradient, eddy PV flux, and LASSO. }
    \label{fig:committor_zfamily}
\end{figure*}

We start by projecting the committor onto each observable at each altitude separately, in hopes of finding particularly salient altitude levels that clarify the role of vertical interactions. The first five rows of Fig. \ref{fig:committor_zfamily} display, for five fields ($U$, $|\Psi|$, $\ov{q'^2}$, $\partial_y\ov q$, and $\ov{v'q'}$) and for a range of altitude levels, the mean and standard deviation of the committor projected onto that field at that altitude. Each altitude has a different range of the CV; for example, because $U$ has a Dirichlet condition at the bottom and a Neumann condition at the top, the lower levels have a much smaller range of variability than the high levels. We also plot the integrated variance, or $L^2$ projection error, at each level in the right-hand column. A low projected committor variance over $U$ at altitude $z_0$ means that the committor is mostly determined by the single observable $U(z_0)$, while a high projected variance indicates significant dependence of $\qp$ on variables other than $U(z_0)$. In order to compare different altitudes and fields as directly as possible, the $L^2$ projection error at each altitude is an average over discrete bins of the observable.

In selecting good CV's, we generally look for a simple, hopefully monotonic, and sensitive relationship with the committor. Of all the candidate fields, $U$ and $\partial_y\ov q$ stand out the most in this respect, being clearly negatively correlated with the forward committor at all altitudes. The associated projection error tends to be greatest in the region $\qp\approx0.5$, as observed before, but interestingly there is a small altitude band around $15-25$ km where its magnitude is minimized. This suggests an optimal altitude for monitoring the committor through zonal wind, giving the most reliable estimate possible for a single state variable. In contrast, the projection of $\qp$ onto $|\Psi|$, displays a large variance across all altitudes. The eddy enstrophy and potential vorticity flux are also rather unhelpful as early warning signs, despite their central role in SSW evolution. For example, the large, positive spikes in heat flux across all altitudes generally occur after the committor $\approx0.5$ threshold has already been crossed. Furthermore, the relationship of $\ov{v'q'}$ with the committor is not smooth. The $\qp<0.5$ region at each altitude is a thin band near zero. 

\begin{figure*}
    \centering
    \includegraphics[trim={0cm 0 0cm 0},clip,width=0.75\linewidth]{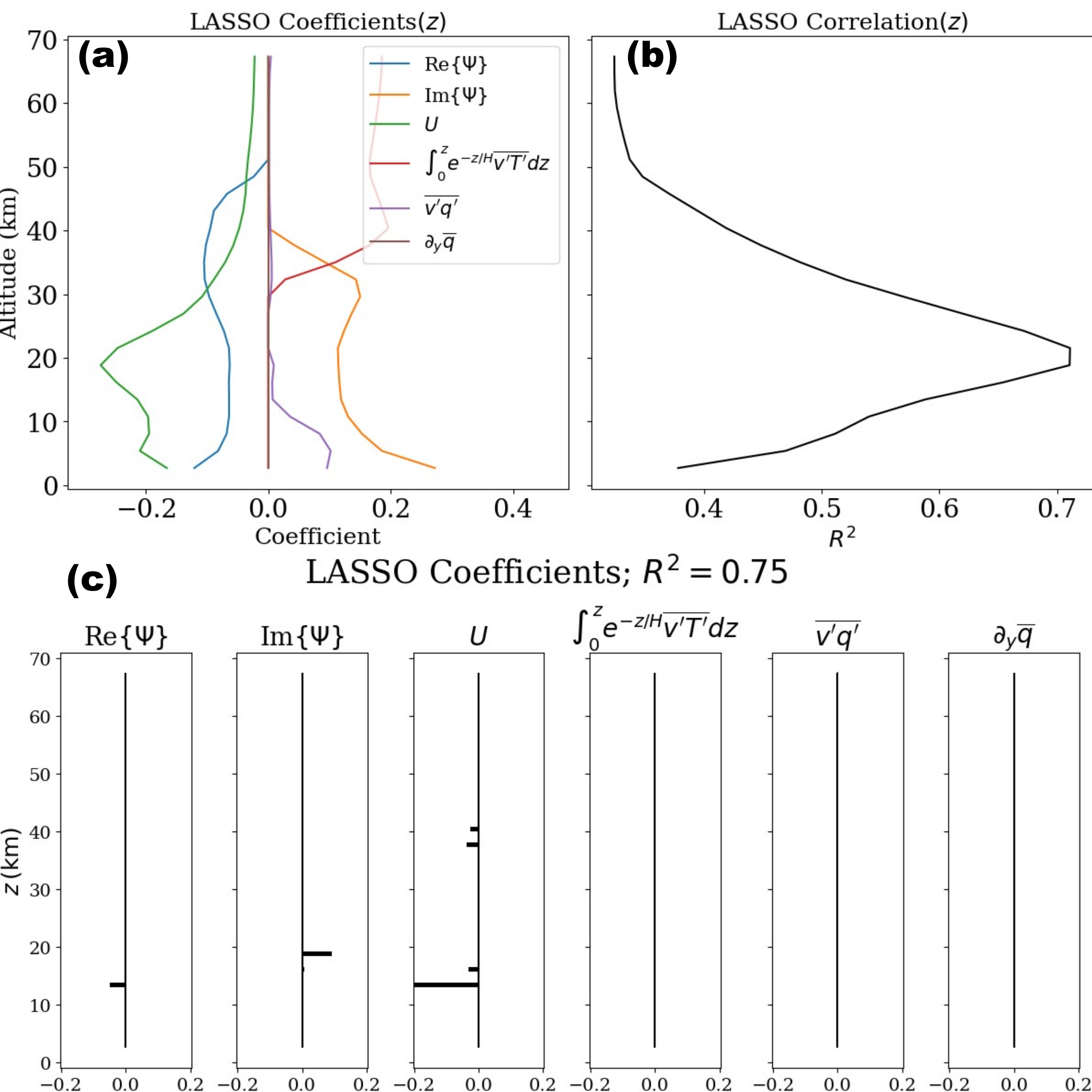}
    \caption{\textbf{Results of LASSO regression of the forward committor with linear and nonlinear input features}. Panel (a) shows the coefficients when $\qp$ is regressed as a function of only the variables at a given altitude, and panel (b) shows the corresponding correlation score. 21.5 km seems the most predictive (where $z\equiv0$ at the tropopause, not the surface). Panel (c) shows the coefficient structure when all altitudes are considered simultaneously. Most of the nonzero coefficients appear between 15-22 km, distinguishing that range as highly relevant for prediction.}
    \label{fig:lasso}
\end{figure*}

The exhaustive CV search in Fig. \ref{fig:committor_zfamily} is visually compelling in favor of some fields and some altitudes over others, but it is not satisfactory as a rigorous comparison. Differences between units and ranges make it difficult to objectively compare the $L^2$ projection error. Furthermore, restricting to one variable at a time is limiting. Accordingly, we also perform a more automated approach to identify salient variables in the form of a generalized linear model for the forward committor, using sparsity-promoting LASSO regression (``Least Absolute Shrinkage and Selection Operator'') due to \cite{Tibshirani1996}, as implemented in the \texttt{scikit-learn} Python package \citep{scikit-learn}. As input features, we use all state variables $\re\{\Psi\},\im\{\Psi\},U$, the integrated heat flux $\int_0^ze^{-z/H}\ov{v'T'}\,dz$, the eddy PV flux $\ov{v'q'}$, and the background PV gradient $\partial_y\ov q$, at all altitudes $z$ simultaneously. The advantage of a sparsity-promoting regression is that it isolates a small number of observables that can accurately approximate the committor in linear combination. Considering that regions close to $A$ and $B$ have low committor uncertainty, we regress only on data points with $\qp\in(0.2,0.8)$, and of those only a subset weighted by $\pi(\vx)\qp(\vx)(1-\qp(\vx))$ to further emphasize the transition region $\qp\approx0.5$. To constrain committor predictions to the range $(0,1)$, we regress on the committor after an inverse-sigmoid transformation, $\ln(\qp/(1-\qp))$. First we do this at each altitude separately, and in Fig. \ref{fig:lasso} (a) we plot the coefficients of each component as a function of altitude. The bottom row of Fig. \ref{fig:committor_zfamily} also displays the committor projected on the height-dependent LASSO predictor.

The height-dependent regression in \ref{fig:lasso}(a) shows each component is salient for some altitude range. In general, $U$ and $\im\{\Psi\}$ dominate as causal variables at low altitudes, while $\re\{\Psi\}$ dominates at high altitudes. The overall prediction quality, as measured by $R^2$ and plotted in Fig. \ref{fig:lasso} (b), is greatest around 21.5 km, consistent with our qualitative observations of Fig. \ref{fig:committor_zfamily}. Note that not all single-altitude slices are sufficient for approximating the committor, even with LASSO regression; in the altitude band $50-60$ km, the LASSO predictor is not monotonic and has a large projected variance, as seen in the bottom row of Fig. \ref{fig:committor_zfamily}. The specific altitude can matter a great deal. But by using all altitudes at once, the committor approximation may be improved further. We thus repeat the LASSO with all altitudes simultaneously and find the sparse coefficient structure shown in \ref{fig:lasso} (c), with a few variables contributing the most, namely the state variables $\Psi$ and $U$ in the altitude range 15-22 km. The nonlinear CVs failed to make any nonzero contribution to LASSO, and this remained stubbornly true for other nonlinear combinations not shown, such as $\ov{v'T'}$. With multiple lines of evidence indicating 21.5 km as an altitude with high predictive value for the forward committor, we can make a strong recommendation for targeting observations here. This conclusion applies only to the Holton-Mass model under these parameters, but the methodology explained above can be applied similarly to models of arbitrary complexity.

We have presented the committor and lead time as ``ideal'' forecasts, especially the committor, which we have devoted considerable effort to approximating in this section. We want to emphasize that $\qp$ and $\lt$ are not competitors to ensemble forecasting; rather, they are two of its most important end results. So far, we have simply advocated including $\qp$ and $\lt$ as quantities of interest. Going forward, however, we do propose an alternative to ensemble forecasting aimed specifically at the committor, lead time, and a wider class of forecasting functions, as they are important enough in their own right to warrant dedicated computation methods. Our approach uses only short simulations, making it highly parallelizable, and shifts the numerical burden from online to offline. Figs. \ref{fig:committor_mfpt_1d}-\ref{fig:lasso} were all generated using the short-simulation algorithm. While the method is not yet optimized and in some cases not competitive with ensemble forecasting, we anticipate such methods will be increasingly favorable with modern trends in computing.

\section{The computational method}
In this section we describe the methodology, which involves some technical results from stochastic processes and measure theory. After describing the theoretical motivation and the numerical pipeline in turn, we demonstrate the method's accuracy and discuss its efficiency compared to straightforward ensemble forecasting.

\subsection{Feynman-Kac formulae}
The forecast functions described above---committors and passage times---can all be derived from general conditional expectations of the form 
\begin{align}
    \label{eq:genexp}
    F(\vx;\lambda)&=\ex_\vx\bigg[G(\bx(\tau))\exp\bigg(\lambda\int_0^\tau\dam(\bx(s))\,ds\bigg)\bigg]
\end{align}
where again the subscript $\vx$ denotes conditioning on $\bx(0)=\vx$; $G,\dam$ are arbitrary known functions over $\R^d$; and $\tau$ is a stopping time, specifically a first-exit time like Eq.~\eqref{eqn:stopping_time} but possibly with $D$ replaced by another set. $\lambda$ is a variable parameter that turns $F$ into a moment-generating function. To see that the forward committor takes on this form, set $G(\vx)=\ind_B(\vx)$, $\lambda=0$ ($\dam$ can be anything), and $\tau=\tau_{A\cup B}$. Then $F(\vx)=\ex_\vx\big[\ind_B(\bx(\tau))\big]=\pr_\vx\{\bx(\tau_{D^c})\in B\}=\qp(\vx)$. For the $\lt$, set $\tau=\tau_B$, $G=\ind_B$, and $\dam=1$. Then 
\begin{align}
    F(\vx;\lambda)&=\ex_\vx\big[\ind_B(\bx(\tau))\exp(\lambda\tau)\big]\\
    \frac1{\qp(\vx)}\pder{}{\lambda}F(\vx;0)&=\frac{\ex_\vx[\tau\ind_B(\bx(\tau))]}{\ex_\vx[\ind_B(\bx(\tau))]} \label{eq:mgf_trick}\\
    &=\lt(\vx).
\end{align}
So we must also be able to differentiate $F$ with respect to $\lambda$.

More generally, the function $G$ is chosen by the user to quantify risk at the terminal time $\tau$; in the case of the forward committor, that risk is binary, with an SSW representing a positive risk and a radiative vortex no risk at all. The function $\dam$ is chosen to quantify the risk accumulated up until time $\tau$, which might be simply an event's duration, but other integrated risks may be of more interest for the application. For example, one could express the total poleward heat flux by setting $\dam=\ov{v'T'}$, or the momentum lost by the vortex by setting $\dam(\vx)=U(\va)-U(\vx)$. Extending~\eqref{eq:mgf_trick}, one can compute not only means but higher moments of such integrals by expressing the risk with $\dam$. Repeated differentiation of $F(\vx;\lambda)$ gives
\begin{align}
    \partial_\lambda^kF(\vx;0)&=\ex_\vx\bigg[G(\bx(\tau))\bigg(\int_0^\tau\dam(\bx(s))\,ds\bigg)^k\bigg] \label{eq:genexp_derivatives}
\end{align}

We choose to focus on expectations of the form \eqref{eq:genexp} in order to take advantage of the Feynman-Kac formula, which represents $F(\vx;\lambda)$ as the solution to a PDE boundary value problem over state space. As PDEs involve local operators, this form is more amenable to solution with short trajectories which don't stray far from their source. The boundary value problem associated with \eqref{eq:genexp} is 
\begin{align}
    \begin{cases}
        (\lcal+\lambda\dam)F(\vx;\lambda)=0 & \vx\in D\\
        F(\vx;\lambda)=G(\vx) & \vx\in D\cm \label{eq:feynman_kac_formula}
    \end{cases}
\end{align}
The domain $D$ here is some combination of $A\cm$ and $B\cm$. The operator $\lcal$ is known as the \emph{infinitesimal generator} of the stochastic process, which acts on functions by pushing expectations forward in time along trajectories:
\begin{align}
    \lcal f(\vx):=\lim_{\Delta t\to0}\frac{\ex_\vx[f(\bx(\Delta t))]-f(\vx)}{\Delta t} \label{eq:generator_definition}
\end{align}
In a diffusion process like the stochastic Holton-Mass model, $\lcal$ is an advection-diffusion partial differential operator which is analogous to a material derivative in fluid mechanics. The generator encapsulates the properties of the stochastic process. In addition to solving boundary value problems \eqref{eq:genexp}, its adjoint $\lcal\str$ provides the Fokker-Planck equation for the stationary density $\pi(\vx)$:
\begin{align}
    \lcal\str\pi(\vx)=0 \label{eq:fokker_planck_equation}
\end{align}
We can also write equations for moments of $F$, as in \eqref{eq:genexp_derivatives}, by differentiating  \eqref{eq:feynman_kac_formula} repeatedly and setting $\lambda=0$:
\begin{align}
    \lcal\big[\partial_\lambda^kF\big](\vx;0)&=-k\dam\partial_\lambda^{k-1}F
    \label{eq:recursion}
\end{align}
This is an application of the Kac Moment Method \citep{Fitzsimmons1999kac}. Note that we never actually have to solve \eqref{eq:feynman_kac_formula} with nonzero $\lambda$. Instead we implement the recursion above. Note that the base case, $k=0$, with $G=\ind_B$ gives $F^+=\qp$, no matter what the risk function $\dam$. In this paper we compute only up to the first moment, $k=1$. Further background regarding stochastic processes and Feynman-Kac formulae can be found in \cite{Karatzas,Oksendal,weinan2019applied}. 

\subsection{Dynamical Galerkin Approximation}
To solve the boundary value problem \eqref{eq:feynman_kac_formula} with $\lambda=0$, we start by following the standard finite element recipe, converting to a variational form and projecting onto a finite basis. First, we homogenize boundary conditions by writing $F(\vx)=\hat F(\vx)+f(\vx)$, where $\hat F$ is a guess function that obeys the boundary condition $\hat F|_{D\cm}=G$, and $f|_{D\cm}=0$. Next, we integrate the equation against any test function $\phi$, weighting the integrand by a density $\mu$ (which is arbitrary for now, but will be specified later):
\begin{align}
    \int_{\R^d}\phi(\vx)\lcal f(\vx)\mu(\vx)\,d\vx&=\int\phi(\vx)(G-\lcal\hat F)(\vx)\mu(\vx)d\vx\nonumber\\
    \ang{\phi,\lcal f}_\mu&=\ang{\phi,G-\lcal\hat F}_\mu \label{eq:weak_formulation}
\end{align}
The test function $\phi$ should live in the same space as $f$, i.e., with homogeneous boundary conditions $\phi(\vx)=0$ for $\vx\in A\cup B$. We refer to the inner products in \eqref{eq:weak_formulation} as being ``with respect to'' the measure (with density) $\mu$. We approximate $f$ by expanding in a finite basis $f(\vx)=\sum_{j=1}^M\xi_j\phi_j(\vx)$ with unknown coefficients $\xi_j$, and enforce that \eqref{eq:weak_formulation} hold for each $\phi_i$. This reduces the problem to a system of linear equations,
\begin{align}
    \sum_{j=1}^M\ang{\phi_i,\lcal\phi_j}_\mu\xi_j&=\ang{\phi_i,G-\lcal\hat F}_\mu && i=1,\hdots,M \label{eq:weak_formulation_basis}
\end{align}
which can be solved with standard numerical linear algebra packages. 

This procedure consists of three crucial subroutines. First, we must construct a set of basis functions $\phi_j$. Second, we have to evaluate the generator's action on them, $\lcal\phi_j$. Third, we have to compute inner products. With standard PDE methods, the basis size would grow exponentially with dimension, quickly rendering the first and third steps intractable. Successful approaches will involve a representation of the solution, $F$, suitable for the high dimensional setting, i.e. representations of the type commonly employed for machine learning tasks. DGA is one such method, whose special twist is to construct a ``data-informed'' basis of reasonable size, evaluate the generator by implementing Eq.~\eqref{eq:generator_definition} with the same data set, and finally evaluate the inner products \eqref{eq:weak_formulation} with a Monte Carlo integral. The data consist of short trajectories launched from all over state space, which the system of linear equations stitches together into a global function estimate. We sketch the procedure here, but for the implementation details we refer to the appendix and to \cite{dga} and \cite{Strahan2020}, where DGA has already been developed for molecular dynamics. 

\vspace{0.5cm}

\noindent\textbf{\underline{Step 1}}: Generate the data, in the format of $N$ initial conditions $\{\bx_n:1\leq n\leq N\}$. Evolve each initial condition forward for a ``lag time'' $\Delta t$ to obtain a set of short trajectories $\{\bx_n(t):0\leq t\leq{\Delta t}, n=1,\hdots,N\}\subset\R^d$. (Lag time is an algorithmic parameter for DGA. It is not to be confused with the forecast time horizon between the prediction and the event of interest in meteorology.) Here and going forward, $\bx_n$ will mean $\bx_n(0)$. The choice of starting points is flexible, but crucial for the efficiency and accuracy of DGA. Because our goal here is to demonstrate interpretable results, we prioritize simplicity and accuracy over efficiency, and defer optimization to later work. We simply draw initial conditions at random from the long control simulation of $5\times10^5$ days, and then generate new short trajectories from those points. We do not sample the points with equal probability, but instead re-weight to get a uniform distribution over the space $(U(30\,\mathrm{km}),|\Psi|(30\,\mathrm{km}))$, within the bounds realized by the control simulation, which are approximately $-30\, \mathrm{m/s}\leq U(30\,\mathrm{km})\leq70$ m/s and $0\,\text{m}^2/\text{s}\leq|\Psi|(30\,\text{km})\leq 2\times10^7\,\text{m}^2/\text{s}$. This sampling procedure, and any other version, implicitly defines a \emph{sampling measure} $\mu$ on state space, where $\mu(\vx)\,d\vx$ is the expected fraction of starting points in the neighborhood $d\vx$ about $\vx$. Sampling points with equal weight from the control run would induce $\mu=\pi$, a very inefficient choice because probability concentrates around the metastable states $\va$ and $\vb$. The re-weighting procedure ensures data coverage of intermediate-wind regions between $A$ and $B$, as well as the large bursts of wave amplitude that characterize the transition pathways. Our main results use $N=5\times10^5$ short trajectories with a lag time of $\Delta t=20$ days, sampled at a frequency of twice per day. This data set is more than needed to get a reasonable committor estimate, but we have sampled generously in order to visualize the functions in high detail. The final section will show the method is robust, capable of reasonably approximating the committor even with an order-of-magnitude reduction in data.

\vspace{0.5cm}

\noindent\textbf{\underline{Step 2}}: Define the basis. The Galerkin method works for any class of basis functions that becomes increasingly expressive as the library grows and becomes capable of estimating any function of interest. However, with a finite truncation, choosing basis functions is a crucial ingredient of DGA, greatly impacting the efficiency and accuracy of the results. In our current study, we restrict to the simplest kind of basis, which consists of indicator functions $\phi_i(x)=\ind_{S_i}(x)$, where $\{S_1,\hdots,S_M\}$ is a disjoint partition of state space. In practice we will construct these sets by clustering the initial data points as described in more detail in Appendix A. This is a common practice in the computational statistical mechanics community for building a Markov State Model (MSM) \citep{Chodera2006longtime,Noe2008,Pande2010,bowman2013introduction,Chodera2014}. MSMs are a dimensionality reduction technique that has also been used in conjuction with analysis of metastable transitions, primarily in protein folding dynamics \citep{Jayachandran2006folding,Noe2009}. MSMs have also been used recently to study garbage patch dynamics in the ocean \citep{Miron2020} as well as complex social dynamics \citep{Helfmann2021statistical}. In \cite{Maiocchi2020unstable}, the authors take an interesting approach to MSMs by clustering points based on proximity to unstable periodic orbits, a potentially useful paradigm for general chaotic weather phenomena \citep{Lucarini2020new}.
DGA can be viewed as an extension of MSMs, though, rather than producing any reduced complexity model,  the explicit goal in DGA is estimating specific functions as in Eq.~\eqref{eq:genexp}.

\vspace{0.5cm}
\noindent\textbf{\underline{Step 3}:} Apply the generator. The forward difference formula
\begin{align}
    \widehat{\lcal\phi}(\bx_n)=\frac{\phi(\bx_n(\Delta t))-\phi(\bx_n)}{\Delta t} 
    \label{eq:forward_difference}
\end{align}
suggested by the definition of the generator \eqref{eq:generator_definition}, results in a systematic bias when $\Delta t$ is finite.  On the other hand, small values of $\Delta t$ lead to large variances in our Monte Carlo estimates of the inner products in \eqref{eq:weak_formulation_basis}.  To resolve these issues we use an integrated form of the Feynman--Kac equations that involves stopping trajectories when they enter $A$ or $B$. 
%and that is exact at any value of $\Delta t$.  
Details are provided in Appendix A.

\vspace{0.5cm}
\noindent\textbf{\underline{Step 4}:} Compute the inner products. The inner products in Eq.~\eqref{eq:weak_formulation_basis} are integrals over high-dimensional state space that are intractable with standard quadrature, but can be approximated using Monte Carlo integration. If $\bx$ is an $\R^d$-valued random variable distributed according to $\mu$, and we have access to random samples $\{\bx_1,\hdots,\bx_N\}$ (which we do), the law of large numbers gives, for any function $g$ with finite expectation,
\begin{align}
    \lim_{N\to\infty}\frac1N\sum_{n=1}^Ng(\bx_n)=\int_{\R^d}g(\vx)\mu(\vx)d\vx
\label{eq:loln}
\end{align}
Setting $g(\vx)=\phi_i(\vx)\lcal\phi_j(\vx)$, the sample average on the left-hand side of \eqref{eq:loln} therefore provides an estimator of $\ang{\phi_i,\lcal\phi_j}_\mu$.
Of course, our approximation uses finite $N$ and nonzero $\Delta t$. A similar sample average approximation can be used to estimate the inner product on the right-hand side of \eqref{eq:weak_formulation_basis}.

These same steps apply to both $\qp$ and $\ex[\tau_B]$, as well as the recursion in \eqref{eq:recursion} for $\lt$. For the Fokker-Planck equation \eqref{eq:fokker_planck_equation}, one extra step is needed to convert an equation with $\lcal\str$ into an equation with $\lcal$. Our procedure for estimating $\pi$ is described in Appendix A.

\vspace{0.5cm}
\noindent\textbf{\underline{Step 5}}: Solve the equation~\eqref{eq:weak_formulation_basis}. With a reasonable basis size $M\lesssim1000$, an $LU$ solver such as in LAPACK via Numpy can handle Eq.~\eqref{eq:weak_formulation_basis}. In the case of the homogeneous system for $w(\vx)$, a $QR$ decomposition can identify the null vector. 

\subsection{DGA fidelity and sensitivity analysis}

\begin{figure*}
    % fidelity
    \centering
    \includegraphics[trim={0cm 0cm 0cm 0},clip,width=0.8\linewidth]{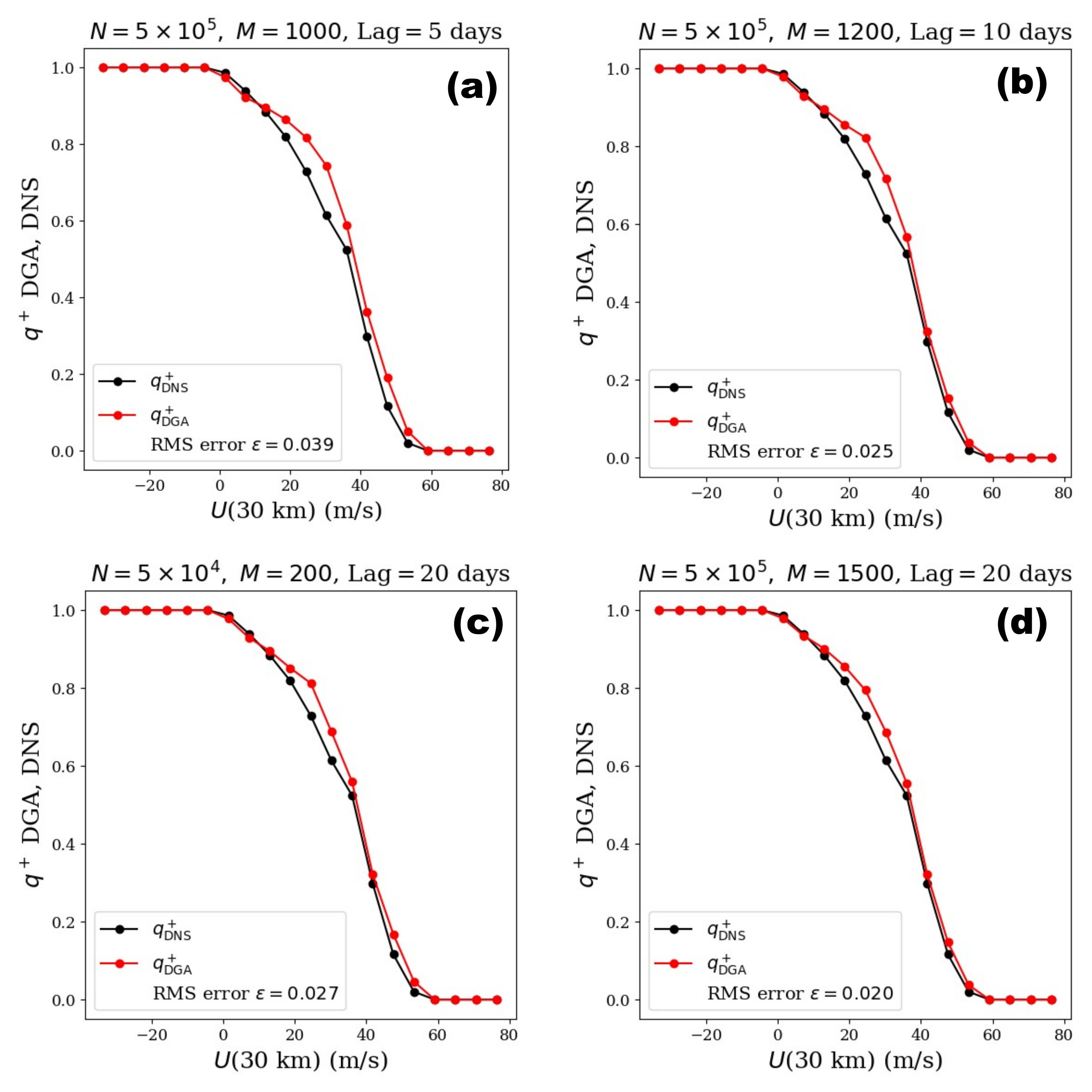}
    \caption{\textbf{Fidelity of DGA.} For several DGA parameter values of $N$ (the number of data points), $M$ (the number of basis functions) and lag time, we plot the committor calculated from DGA and DNS (from the long control simulation), both as a function of $U$(30 km). The mean-square difference $\eps$ in the legend is used as a global error estimate for DGA.}
    \label{fig:committor_fidelity}
\end{figure*}

To illustrate the effect of parameter choices on performance, we present here a simple sensitivity analysis. Fig. \ref{fig:committor_fidelity} verifies the numerical accuracy and convergence of DGA by plotting the committor as a function of $U$(30 km), estimated both with DNS and DGA, for various DGA parameters. The red curves $\qp_\text{DGA}(U(30\,\text{km}))$ are calculated by projecting the committor as in Fig. \ref{fig:committor_mfpt_1d}(a), while the black curve $\qp_\text{DNS}(U(30\,\mathrm{km}))$ is an empirical committor estimate equal to the fraction of control simulation points seen at a particular value of $U(30\,\mathrm{km})$ that next hit $B$. 

In panels (a), (b), and (d), the lag time $\Delta t$ increases from 5 to 10 to 20 days while the number of short trajectories stays fixed at $N=5\times10^5$. Panel (c) has a long lag of 20 days, but a small data set of $N=5\times10^4$, allowing us to see the tradeoff between $N$ and $\Delta t$. The basis size $M$ is chosen heuristically as large as possible within reason for the clustering algorithm (see Appendix A). While DGA tends to systematically overestimate $\qp$ relative to $\qp_\text{DNS}$ in the mid-range of $U$, it seems to approach the empirical estimate as the data size and lag time increase. Each plot also displays the root-mean-square deviation between the two estimators over this subspace, $\eps=\sqrt{\big\langle(\qp_\text{DGA}-\qp_\text{DNS})^2\big\rangle_\pi}$. Within this regime, it seems that increasing the lag time has a greater impact on the deviation than increasing the number of data points. Panels (b) and (c) have approximately the same deviation $\eps$, but (c) uses only one fifth the data, measured by total simulation time. On the other hand, more short trajectories can be parallelized more readily than fewer long trajectories, and the optimal choice will depend on computing resources. 

It is natural to ask whether our short trajectory based approach is more efficient than DNS in which many independent ``long'' trajectories are launched from a single initial condition $\vx$ and the committor probability $\qp(\vx)$ (or another forecast) is estimated directly. For a single value of $\vx$ for which $\qp(\vx)$ is not very small (so that a non-negligible fraction of trajectories reach $B$ before $A$) and for which the lead time $\lt(\vx)$ is not too large (so that trajectories reaching $B$ do so without requiring long integration times), DNS will undoubtedly be more efficient. This is often the situation in real-time weather forecasting. However, a key feature of our approach is that it simultaneously estimates forecasts at all values of $\vx$, allowing the subsequent analysis of those functions that has been the focus of much of this article. Global knowledge of the committor and lead time is more pertinent for oft-repeated forecasts, for long-term risk assessment of extreme event climatology, and for targeting observations optimally. Building accurate estimators in all of state space by DNS would be extremely costly even for the reduced complexity model studied here. 

\section{Conclusion}
Forecasting rare events is, by the very nature of rare events, an extremely difficult computational task, and one of science's most pressing challenges. We have described a computational framework, a dynamical Galerkin approximation to the Feynman-Kac equations, that combines the minimalistic philosophy of dimensionality reduction with the fidelity of high-resolution models. We identify a set of reduced coordinates, the committor probability and expected lead time, that provide the  essential information that large ensemble forecasts hope to compute. DGA uses relatively short simulations of the full model to estimate these quantities of interest, allowing for prediction on much longer timescales than that of the simulation. In its focus on directly estimating statistics of interest, DGA differs from previous reduced-order modeling methods that attempt to capture general qualities of the system, including both physics-based models \citep{Lorenz1963,Charney1979,Legras1985,Crommelin2003,Timmerman2003,ruz} and more recent data-driven models making use of machine learning \citep{nlsa,giannakis_kolchinskaya_krasnov_schumacher_2018,Berry2015,Sabeerali2017,Majda2018strategies,Wan2018data-assisted,Bolton2019,Chattopadhyay2020,Chen2020predicting,Kashinath2021physics,Chattopadhyay2021towards}.

We have shown numerical results in the context of a stochastically forced Holton-Mass model with 75 degrees of freedom, which points to the method's promise for forecasting. By systematically evaluating many model variables for their utility in predicting the fate of the vortex, we have identified some salient physical descriptions of early warning signs. We have furthermore examined the relationship between probability and lead time for a given rare event, a powerful pairing for assessing predictability and preparing for extreme weather. Our results suggest that the slow evolution of vortex preconditioning is an important source of predictability. In particular, the zonal wind and streamfunction in the range of 10-20 km above the tropopause seems to be optimal among a large class of dynamically motivated observables. 

Beyond the problem of real-time weather forecasting, it is also important to assess the climatology, i.e., long-term frequency, intensity, and other characteristics of rare events. For this goal as well, our methodology offers advantages over large ensemble simulations, which are currently the most detailed source of data \citep[e.g.,][]{Schaller2018}. The committor and lead time are ingredients in a larger framework called Transition Path Theory (TPT) for describing rare transition events \emph{at steady state}, meaning average properties over long timescales. TPT describes not only the future evolution from an initial condition ($\vx\to B$), but the ensemble of full vortex breakdown events ($A\to B$), and how they differ from restoration events ($B\to A$). In principle, interrogating the ensemble of transition paths requires direct simulation of the system long enough to observe many transition events. However, using TPT, quantities computable by our framework can be combined to yield key statistics describing the ensemble of transition paths \citep{tpt_simple_examples,tpt_mjp,pathfinding,Weinan2006,Finkel2020}. In a following paper we will apply the same short-trajectory forecasting approach together with TPT to compute transition path statistics such as return times and extract insight about physical mechanisms of the transition process.

Scaling our approach up to state-of-the-art weather and climate models will require significant further development.  In particular, a completely new procedure for generating trajectory initial conditions will need to be introduced. Generation of a trajectory long enough to thoroughly sample transitions will not be practical for more complicated models.  One promising alternative is launching many trajectories in parallel and selectively replicating those that explore new regions of state space, especially transition regions. Such an approach could build on exciting progress over the last decade in  targeted rare event simulation schemes \citep{Hoffman2006response,Weare2009,Bouchet2011,Bouchet2014_Langevin,VandenEijnden2013,Chen2014mcmc,bouchet,Farazmand2017variational,rogue,Mohamad2018sequential,Dematteis2019experimental,webber,Bouchet2019_nucleations,Bouchet2019,Plotkin2019,simonnet2020multistability,Ragone2020averaged,Sapsis2021statistics}. A potential challenge here is that GCMs may not be set up for short simulations that start and stop frequently. For this reason, it may be sensible to use longer lag times and a sliding window to define short trajectories. Furthermore, the communication overhead required for adaptive sampling with GCMs would impose additional costs. We have deferred the sampling problem to future work, acknowledging that this step is crucial to make DGA competitive. The utility of committor and lead time, however, is independent of the method for computing them.

Defining the source of stochasticity is also an important step that varies between models. Explicitly stochastic parameterization \citep[e.g.,][]{Berner2009skebs,Portamana2014stochastic} will automatically lead to a spread in the short-trajectory ensemble, but in deterministic models, uncertainty will arise from perturbing the initial conditions. This may require special care depending on the model.

Another area of algorithmic improvement is selecting a basis expansion of the forecast functions. In upcoming work we will explore more flexible representations using kernel methods and neural networks. The solution of high-dimensional PDEs is an active research area that is making innovative use of machine learning, particularly in the fields of computational chemistry, quantum mechanics, and fluid dynamics \citep[e.g.,][]{Carleo2017,Han2018,Khoo2018,Li2020,Mardt2018vampnets,Li2019computing,Raissi2019pinn,Lorpaiboon2020ivac,rotskoff2020learning}. Similar approaches may hold great potential for understanding predictability in atmospheric science.

%% In all cases, if there is only one entry of this type within
%% the higher level heading, use the star form: 
%%
% \section{Section title}
% \subsection*{subsection}
% text...
% \section{Section title}

%vs

% \section{Section title}
% \subsection{subsection one}
% text...
% \subsection{subsection two}
% \section{Section title}

%%%
% \section{First primary heading}

% \subsection{First secondary heading}

% \subsubsection{First tertiary heading}

% \paragraph{First quaternary heading}

%%%%%%%%%%%%%%%%%%%%%%%%%%%%%%%%%%%%%%%%%%%%%%%%%%%%%%%%%%%%%%%%%%%%%
% ACKNOWLEDGMENTS
%%%%%%%%%%%%%%%%%%%%%%%%%%%%%%%%%%%%%%%%%%%%%%%%%%%%%%%%%%%%%%%%%%%%%
\acknowledgments
J.F. is supported by the U.S. Department of Energy, Office
of Science, Office of Advanced Scientific Computing Research, Department of Energy Computational
Science Graduate Fellowship under Award Number DE-SC0019323.\footnote{This report was prepared as an account of work sponsored by an agency of the United
States Government. Neither the United States Government nor any agency thereof, nor any of their
employees, makes any warranty, express or implied, or assumes any legal liability or responsibility for the
accuracy, completeness, or usefulness of any information, apparatus, product, or process disclosed, or
represents that its use would not infringe privately owned rights. Reference herein to any specific
commercial product, process, or service by trade name, trademark, manufacturer, or otherwise does not
necessarily constitute or imply its endorsement, recommendation, or favoring by the United States
Government or any agency thereof. The views and opinions of authors expressed herein do not
necessarily state or reflect those of the United States Government or any agency thereof.}  
R.J.W. was supported during this project by 
New York University's Dean's Dissertation Fellowship and by
the Research Training Group in Modeling and Simulation funded by the National Science Foundation via grant RTG/DMS-1646339.
E.P.G. acknowledges support from the U.~S.~National Science Foundation through grant AGS-1852727. This work was partially supported by the NASA Astrobiology Program, grant No.~80NSSC18K0829 and benefited from participation in the NASA
Nexus for Exoplanet Systems Science research coordination network. J.W. acknowledges support from the Advanced Scientific Computing Research Program within the DOE Office of Science through award DE-SC0020427. The computations in the paper were done on the high-performance computing clusters at New York University and the Research Computing Center at the University of Chicago.

We extend special thanks to Thomas Birner, Pedram Hassanzadeh, and an anonymous reviewer from \emph{Monthly Weather Review}, who provided invaluable feedback on both the technical and high-level aspects of the manuscript. Their insight has helped us to sharpen and clarify the message. Freddy Bouchet also offered constructive feedback. We thank John Strahan, Aaron Dinner, and Chatipat Lorpaiboon for helpful methodological advice. Mary Silber, Noboru Nakamura, and Richard Kleeman offered invaluable scientific insight. J.F. benefitted from many helpful discussions with Anya Katsevich. 

%%%%%%%%%%%%%%%%%%%%%%%%%%%%%%%%%%%%%%%%%%%%%%%%%%%%%%%%%%%%%%%%%%%%%
% DATA AVAILABILITY STATEMENT
%%%%%%%%%%%%%%%%%%%%%%%%%%%%%%%%%%%%%%%%%%%%%%%%%%%%%%%%%%%%%%%%%%%%%
% 
%
\datastatement
The code for simulating the model, performing DGA, and producing plots is publicly available in the SHORT Github repository, ``Solving for Harbingers Of Rare Transitions'', at https://github.com/justinfocus12/SHORT. J.F. is happy to provide further guidance upon request. 

%%%%%%%%%%%%%%%%%%%%%%%%%%%%%%%%%%%%%%%%%%%%%%%%%%%%%%%%%%%%%%%%%%%%%
% APPENDIXES
%%%%%%%%%%%%%%%%%%%%%%%%%%%%%%%%%%%%%%%%%%%%%%%%%%%%%%%%%%%%%%%%%%%%%

\appendix[A]
\appendixtitle{Feynman-Kac formula and DGA}
In this section we spell out the DGA procedure in more detail than the main text, explaining the variants that get us to the more intricate conditional expectations. The theoretical background can be found in, e.g., \cite{Karatzas,Oksendal,weinan2019applied}. 
Let $\bx(t)$ be a time-homogeneous stochastic process with continuous sample paths in $\R^d$. Associated to this process is the infinitesimal generator, $\lcal$, which acts on functions of state space (also called ``observable'' functions) by evolving their expectation forward in time:
\begin{align}
    \lcal f(\vx)&=\lim_{\Delta t\to0}\frac{\ex_\vx[f(\bx(\Delta t))]-f(\vx)}{\Delta t}
    \label{eq:generator_definition2}
\end{align}
where $\ex_\vx[\cdot]:=\ex[\cdot|\bx(0)=\vx]$. It can be shown that under the above assumptions on $\bx$, the It\^o chain rule gives
\begin{align}
    df(\bx(t))&=\lcal f(\bx(t))\,dt+d\mathbf{M}(t)
\end{align}
where $\mathbf{M}(t)$ is a martingale. More concretely, in this paper, $\bx(t)$ is an It\^o diffusion obeying the stochastic differential equation
\begin{align}
    \bx(t)=\bx(0)&+\int_0^tb(\bx(s))\,ds\\
    &+\int_0^t\bm\sigma(\bx(s))\,d\mathbf{W}(s) \nonumber
\end{align}
with infinitesimal generator and martingale terms
\begin{align}
    \lcal f(\vx)&=\sum_{i=1}^db_i(\vx)\pder{f(\vx)}{\vx_i}\\
    &\hspace{0.2cm}+\sum_{i=1}^d\sum_{j=1}^d\frac12\big[\sigma(\vx)\sigma(\vx)\tr\big]_{ij}\frac{\partial^2f(\vx)}{\partial\vx_i\partial\vx_j} \nonumber\\ 
    d\mathbf{M}(t)&=\sum_{i=1}^d\pder{f(\vx)}{\vx_i}\sigma_{ij}(\vx)d\mathbf{W}_j(t)
\end{align}
The key forecasting quantities in this paper are of the form ~\eqref{eq:genexp} and can be solved with ~\eqref{eq:feynman_kac_formula}, a linear equation involving the generator. We now lay out a brief derivation of the Feynman-Kac formula and our numerical discretization, roughly following \cite{weinan2019applied}.

\subsection{Feynman-Kac formula}
Let $D$ be a domain in $\R^d$ (for example, $(A\cup B)^c$) and $\tau_{D\cm}=\min\{t\geq0:\bx(t)\notin D\}$ be the first exit time from this domain starting at time zero. This is a random variable which depends on the starting condition $\vx\in D$. Let $G:\partial D\to\R$ be a boundary condition, $\dam:D\to\R$ a source term, and $\dam:D\to\R$ a term to represent accumulated risk. We seek a PDE for the conditional expectation from ~\eqref{eq:genexp}:
\begin{align}
    F(\vx)=\ex_\vx\bigg[&G(\bx(\tau))\exp\bigg(\lambda\int_0^{\tau}\dam(\bx(s))\,ds\bigg)\bigg]  \label{eq:genexp2}
\end{align}
where $\ex_\vx[\cdot]=\ex[\cdot|\bx(0)=\vx]$. To derive the PDE ~\eqref{eq:feynman_kac_formula}, consider the following stochastic process:
\begin{align}
    Z(t)&=F(\bx(t))Y(t)
\end{align}
where $Y(t):=\exp\big(\lambda\int_0^t\dam(\bx(s))\,ds\big)$. It\^o's lemma gives us that $dY(t)=\lambda\dam(\bx(t))Y(t)\,dt$. Hence, applying the product rule to $Z(t)$,
\begin{align}
    dZ(t)&=dF(\bx(t))Y(t)+F(\bx(t))\,dY(t)\label{eqn:chainrule}\\
    &=\lcal F(\bx(t))Y(t)\,dt+d\mathbf{M}(t)Y(t)\\
    &\hspace{0.2cm}+\lambda F(\bx(t))\dam(\bx(t))Y(t)\,dt \nonumber\\
    &=\big[\lcal F+\lambda\dam F\big](\bx(t))Y(t)\,dt+Y(t)d\mathbf{M}(t)
\end{align}
where in \eqref{eqn:chainrule} we have left out the quadratic cross-variation of $F(\bx(t))$ and $Y(t)$ because $Y$ has finite variation. If the bracketed term $(\lcal+\lambda\dam(\vx))F(\vx)=0$ for all $\vx$, then $Z(t)$ is a martingale and it follows that 
\begin{align} 
    Z(0)&=\ex_\vx[Z(t)]\\
    F(\vx)&=\ex_\vx\bigg[F(\bx(t))\exp\bigg(\lambda\int_0^t\dam(\bx(s))\,ds\bigg)\bigg]
\end{align}
Finally, the formula still holds if we substitute a stopping time for $t$. By choosing $\tau$, the first exit time from $D$, the $F(\bx(t))$ inside the brackets becomes its boundary value $G(\bx(\tau))$. Thus $F(\vx)$ as defined in ~\eqref{eq:genexp2} also solves the PDE boundary value problem~\eqref{eq:feynman_kac_formula}:
\begin{align}
    \begin{cases}
        (\lcal+\lambda\dam(\vx))F(\vx;\lambda)=0 & \vx\in D\\
        F(\vx;\lambda)=G(\vx) & \vx\in D\cm
        \label{eq:feynman_kac_formula2}
    \end{cases}
\end{align}
where we have inserted the additional dependence of $F$ on $\lambda$ in order to lead directly to the recursive formulas~\eqref{eq:mgf_trick} and~\eqref{eq:recursion}.

\subsection{Dynkin's formula and finite lag time}
We have presented ~\eqref{eq:forward_difference} as a mathematically concise approximation to the generator. In practice, we achieve better numerical stability integrating the generator~\eqref{eq:generator_definition2} to a finite lag time $\Delta t$, following \cite{Strahan2020}. The theorem that allows this is called Dynkin's formula \citep[e.g.,][]{Oksendal}, which states that for any suitable function $f:\R^d\to\R$ and a stopping time $\theta$ (not to be confused with CV coordinates),
\begin{align}
    \ex_\vx[f(\bx(\theta))]=f(\vx)+\ex_\vx\bigg[\int_0^\theta\lcal f(\bx(t))\,dt\bigg].
    \label{eq:dynkin_formula}
\end{align}
The left-hand side, $\ex_\vx[f(\bx(\theta))]$, is known as the \emph{transition operator} $\tcal^\theta f(\vx)$, a finite-time version of the generator. Note that this is a deterministic operator despite $\theta$ being a random variable, because by definition $\tcal^\theta$ only has $\theta$ inside of expectations. We can apply Dynkin's formula to~\eqref{eq:feynman_kac_formula2} \emph{before} numerical approximation, setting $\theta=\min(\Delta t,\tau)$. That is, the short trajectory $\{\bx(t):0\leq t\leq\Delta t=20\text{ days}\}$ is stopped early if it exits the domain $D$ before $\Delta t$. Applying Dynkin's formula to $F(\vx;\lambda)$, we find
\begin{align}
    \ex_\vx[F(\bx(\theta))]&=F(\vx)+\ex_\vx\bigg[\int_0^\theta\lcal F(\bx(t))\,dt\bigg] \nonumber\\
    &=F(\vx)-\lambda\ex_\vx\bigg[\int_0^\theta\dam(\bx(t))F(\bx(t))\,dt\bigg] \nonumber\\
    \tcal^\theta F(\vx)&=F(\vx)-\lambda\ical^\theta[\dam F](\vx)
    \label{eq:feynman_kac_formula_integrated}
\end{align}
where $\ical^\theta$ is shorthand notation for the integral operator on the right. Eq.~\eqref{eq:feynman_kac_formula_integrated}, along with the boundary conditions $F|_{D\cm}=G|_{D\cm}$, gives us a linear equation for $F(\vx)$ that can be solved by DGA. As outlined in Section 5, we write $F=\hat F+f$, where $\hat F$ obeys the boundary conditions and $f$ obeys
\begin{align}
    (\tcal^\theta-1)f(\vx)&+\lambda\ical^\theta[\dam f](\vx)=\\
    &-(\tcal^\theta-1)\hat F(\vx)-\lambda\ical^\theta[\dam\hat F](\vx) \nonumber
\end{align}
We then expand $f=\sum_{j=1}^M\xi_j\phi_j(\vx)$ with basis functions $\{\phi_j\}$ that are zero on $D\cm$, and take $\mu$-weighted inner products with $\phi_i$ on both sides to obtain
\begin{align}
    \sum_{j=1}^M\xi_j\Big(\big\langle&\phi_i,(\tcal^\theta-1)\phi_j\big\rangle_\mu+\lambda\big\langle\phi_i,\ical^\theta[\dam\phi_j]\big\rangle_\mu\Big)=\nonumber\\
    &-\big\langle\phi_i,(\tcal^\theta-1)\hat F\big\rangle_\mu-\lambda\big\langle\phi_i,\ical^\theta[\dam\hat F]\big\rangle_\mu
    \label{eq:dynkin_discretized}
\end{align}

Finally, the inner products can be estimated with short trajectories using~\eqref{eq:loln}. For two functions $\phi$ and $\psi$, the first left-hand side inner product is approximately
\begin{align}
    \big\langle\phi,(\tcal^\theta-1)\psi\big\rangle_\mu&\approx\frac1N\sum_{n=1}^N\phi(\bx_n)\big[\psi(\bx_n(\theta_n))-\psi(\bx_n)\big]
    \label{eq:generator_matrix_elements}
\end{align}
where $\theta_n$ is the sampled first-exit time of the $n$th trajectory, or $\Delta t$ if it never exits. The second left-hand side inner product is approximately
\begin{align}
    \big\langle\phi,&\ical^\theta[\dam\psi]\big\rangle_\mu\approx \nonumber\\
    &\frac1N\sum_{n=1}^N\phi(\bx_n)\int_0^{\theta_n}\dam(\bx_n(t))\psi(\bx_n(t))\,dt
\end{align}
where the time integral on the right is computed with the trapezoid rule on the trajectory, which is sampled every $0.5$ days. The error from numerical quadrature is likely small compared to the error from basis set construction, but higher-order integration methods do merit further investigation.

Given a fixed $\dam$ and $G$, and with the inner products in hand, we now have~\eqref{eq:dynkin_discretized} as a family of matrix equations with $\lambda$ a continuous parameter:
\begin{align}
    (P+\lambda Q)\bm\xi(\lambda)=\mathbf{v}+\lambda\mathbf{r}.
\end{align}
We can then differentiate in $\lambda$ and evaluate at $\lambda=0$ to obtain a ready-to-solve discretization of the recursion~\eqref{eq:recursion}:
\begin{align}
    P\bm\xi(0)&=\mathbf{v}\\
    P\bm\xi'(0)&=\mathbf{r}-Q\xi(0)\\
    P\bm\xi\s{k}(0)&=-kQ\bm\xi\s{k-1}(0)\text{ for }k\geq2
\end{align}
where the $k$'th derivative $\bm\xi\s{k}(0)$ is the coefficient expansion in the basis $\{\phi_j\}$ of the $k$'th moment from~\eqref{eq:genexp_derivatives}:
\begin{align}
    \partial_\lambda^kF(\vx;0)&=\ex_\vx\bigg[G(\bx(\tau))\bigg(\lambda\int_0^\tau\dam(\bx(s))\,ds\bigg)^k\bigg]
\end{align}

\subsection{Change of measure}
We now specify how to compute the change of measure from $\mu$ (the sampling distribution) to $\pi$ (the steady-state distribution), using an adjoint version of the Feynman-Kac formula. Each of the basis functions $\phi_i$ has an expectation at time zero with respect to the steady state distribution: $\ex_{\bx(0)\sim\pi}[\phi_i(\bx(0))]=\int\phi_i(\vx)\pi(\vx)\,d\vx$. Evolving the dynamics from $0$ to $\Delta t$ induces another expectation: $\ex_{\bx(0)\sim\pi}[\phi_i(\bx(\Delta t))]=\int\tcal^{\Delta t}\phi_i(\vx)\pi(d\vx)$. $\pi$ is the \emph{invariant} distribution, which means that these two integrals are equal:
\begin{align}
    \int(\tcal^{\Delta t}-1)\phi_i(\vx)\pi(\vx)d\vx=0.
\end{align}
Furthermore, with a change of measure they can be rewritten with respect to the sampling measure $\mu$ instead of $\pi$, so 
\begin{align}
    \int(\tcal^{\Delta t}-1)\phi_i(\vx)\der{\pi}{\mu}(\vx)\mu(\vx)\,d\vx=0
\end{align}
The change of measure $\der\pi\mu(\vx)$, which we abbreviate $w(\vx)$, is yet another unknown function which we expand in the basis as $w(\vx)=\sum_j\xi_j\phi_j(\vx)$. Putting this into the integral and using Monte Carlo, we cast the coefficients $\xi_j$ as the solution to a null eigenvector problem:
\begin{align}
    0&=\int(\tcal^{\Delta t}-1)\phi_i(\vx)\sum_{j=1}^M\xi_j\phi_j(\vx)\mu(d\vx)\\
    &\approx\sum_{j=1}^M\xi_j\sum_{n=1}^N\big[\phi_i(\bx_n(\Delta t))-\phi_i(\bx_n)\big]\phi_j(\bx_n)
    \label{eq:fokker_planck_discretized}
\end{align}
This last equation is simply the Fokker-Planck equation, $\lcal\str\pi=0$, in weak form and integrated in time using Dynkin's formula. Note that the matrix elements in~\eqref{eq:fokker_planck_discretized} are the transpose of those in~\eqref{eq:generator_matrix_elements}.

\subsection{DGA details}
We will provide more details here on our particular construction of basis functions. The partition $\{S_1,\hdots,S_M\}$ to build the basis function library $\phi_j(\vx)=\ind_{S_j}(\vx)$, $n=1,\hdots,N$ should be chosen with a number of considerations in mind. The partition elements should be small enough to accurately represent the functions they are used to approximate, but large enough to contain sufficient data to robustly estimate transition probabilities. We form these sets by a hierarchical modification of $K$-means clustering on the initial points $\{\bx_n\}_{n=1}^N$. $K$-means is a robust method that can incorporate new samples by simply identifying the closest centroid, and is commonly used in molecular dynamics \citep{Pande2010}. However, straightforward application of $K$-means, as implemented in the \texttt{scikit-learn} software \citep{scikit-learn}, can produce a very imbalanced cluster size distribution, even with empty clusters. This leads to unwanted singularities in the constructed Markov matrix. To avoid this problem we cluster hierarchically, starting with a coarse clustering of all points and iteratively refining the larger clusters, at every stage enforcing a minimum cluster size of five points, until we have the desired number of clusters ($M$). After clustering on the initial points $\{\bx_n\}$, the other points $\{\bx_n(t),0<t\leq\Delta t\}$ are placed into clusters using an address tree produced by the $K$-means cluster hierarchy. For boundary value problems with a domain $D$ and boundary $D\cm$, we need only cluster points in $D$, since the basis should be homogeneous. The total number of clusters should scale with data set. In our main results with $N=5\times10^5$, we found $M=1500$ to be enough basis functions to resolve some of the finer details in the structure of the forecast functions, but not so many as to require an unmanageably deep address tree, which manifests in dramatic slowdown past a certain threshold. At this point, the cluster number is still a manually tuned hyperparameter.

Because the committor and lead time obey Dirichlet boundary conditions on $A\cup B$, the basis funtions used to construct them should be zero on $A\cup B$, meaning only data points $\bx_n\notin A\cup B$ should be used to produce the clusters. On the other hand, the steady state distribution has no boundary condition to satisfy, only a global normalization condition. Hence, the basis for the change of measure $w$ must be different from the basis for $\qp$ and $\lt$, with its clusters including all data points in $A\cup B$. Furthermore, the basis must be chosen so that the matrix $\ang{(\tcal^{\Delta t}-1)\phi_i,\phi_j}$ has a nontrivial null space; this is guaranteed by the indicator basis set we use, but can otherwise be guaranteed by including a constant function in the basis. 

The use of an indicator basis follows the Markov State Modeling literature \citep[e.g.,]{Chodera2006longtime,Pande2010}, which has the advantage of simplicity and robustness. In particular, the discretization of $\tcal^\theta-1$ is a properly normalized stochastic matrix (with nonnegative entries and rows summing to 1), which guarantees the maximum principle $0\leq\qp(\vx)\leq1$ and $0\leq w(\vx)$ for all data points $\vx$. However, alternative basis sets have been shown to be promising, perhaps with much less data. \cite{dga} used diffusion maps, while \cite{Strahan2020} used a PCA-like procedure to construct the basis. More generally, there is no requirement to use a linear Galerkin method to solve the Feynman-Kac formulae. More flexible functional forms may have an important role to play as well. In the low-data regime, some preliminary experiments have suggested that Gaussian process regression (GPR) is a useful way to constrain the committor estimate with a prior, following the framework in \cite{Bilionis2016gpr} to solve PDEs with Gaussian processes. As mentioned in the conclusion, there is rapidly growing interest in the use of artificial neural networks to solve PDEs. As with many novel methods, however, DGA is likely to work best on new applications when its simplest form is applied first. This will be our approach in coming experiments on more complex models. 

%
% Use \appendix if there is only one appendix.
%\appendix

% Use \appendix[A], \appendix[B], if you have multiple appendixes.
%\appendix[A]

%% Appendix title is necessary! For appendix title:
%\appendixtitle{}

%%% Appendix section numbering (note, skip \section and begin with \subsection)
% \subsection{First primary heading}

% \subsubsection{First secondary heading}

% \paragraph{First tertiary heading}

%% Important!
%\appendcaption{<appendix letter and number>}{<caption>} 
%must be used for figures and tables in appendixes, e.g.,
%
%\begin{figure}
%\noindent\includegraphics[width=19pc,angle=0]{figure01.pdf}\\
%\appendcaption{A1}{Caption here.}
%\end{figure}
%
% All appendix figures/tables should be placed in order AFTER the main figures/tables, i.e., tables, appendix tables, figures, appendix figures.
%
%%%%%%%%%%%%%%%%%%%%%%%%%%%%%%%%%%%%%%%%%%%%%%%%%%%%%%%%%%%%%%%%%%%%%
% REFERENCES
%%%%%%%%%%%%%%%%%%%%%%%%%%%%%%%%%%%%%%%%%%%%%%%%%%%%%%%%%%%%%%%%%%%%%
% Make your BibTeX bibliography by using these commands:
\bibliographystyle{ametsoc2014_arxiv}
\bibliography{references}

\begin{thebibliography}{98}
\providecommand{\natexlab}[1]{#1}
\providecommand{\url}[1]{\texttt{#1}}
\renewcommand{\UrlFont}{\rmfamily}
\providecommand{\urlprefix}{URL }
\expandafter\ifx\csname urlstyle\endcsname\relax
  \providecommand{\doi}[1]{doi:\discretionary{}{}{}#1}\else
  \providecommand{\doi}{doi:\discretionary{}{}{}\begingroup
  \urlstyle{rm}\Url}\fi
\providecommand{\eprint}[2][]{\url{#2}}

\bibitem[{Berner et~al.(2009)Berner, Shutts, Leutbecher,, and
  Palmer}]{Berner2009skebs}
Berner, J., G.~J. Shutts, M.~Leutbecher, and T.~N. Palmer, 2009: A spectral
  stochastic kinetic energy backscatter scheme and its impact on flow-dependent
  predictability in the ecmwf ensemble prediction system. \textit{Journal of
  the Atmospheric Sciences}, \textbf{66~(3)}, 603 -- 626,
  \doi{10.1175/2008JAS2677.1},
  \urlprefix\url{https://journals.ametsoc.org/view/journals/atsc/66/3/2008jas2677.1.xml}.

\bibitem[{Berry et~al.(2013)Berry, Cressman, Gregurić-Ferenček,, and
  Sauer}]{Berry2013timescale}
Berry, T., J.~R. Cressman, Z.~Gregurić-Ferenček, and T.~Sauer, 2013:
  Time-scale separation from diffusion-mapped delay coordinates. \textit{SIAM
  Journal on Applied Dynamical Systems}, \textbf{12~(2)}, 618--649,
  \doi{10.1137/12088183X}, \urlprefix\url{https://doi.org/10.1137/12088183X},
  \eprint{https://doi.org/10.1137/12088183X}.

\bibitem[{Berry et~al.(2015)Berry, Giannakis,, and Harlim}]{Berry2015}
Berry, T., D.~Giannakis, and J.~Harlim, 2015: Nonparametric forecasting of
  low-dimensional dynamical systems. \textit{Phys. Rev. E}, \textbf{91},
  032\,915, \doi{10.1103/PhysRevE.91.032915}.

\bibitem[{Bilionis(2016)}]{Bilionis2016gpr}
Bilionis, I., 2016: Probabilistic solvers for partial differential equations.
  \textit{arXiv: Probability}.

\bibitem[{Binzel(2000)}]{Binzel2000torino}
Binzel, R.~P., 2000: The torino impact hazard scale. \textit{Planetary and
  Space Science}, \textbf{48~(4)}, 297--303,
  \doi{https://doi.org/10.1016/S0032-0633(00)00006-4},
  \urlprefix\url{https://www.sciencedirect.com/science/article/pii/S0032063300000064}.

\bibitem[{Birner and Williams(2008)Birner, and Williams}]{Birner2008}
Birner, T., and P.~D. Williams, 2008: Sudden stratospheric warmings as
  noise-induced transitions. \textit{Journal of the Atmospheric Sciences},
  \textbf{65~(10)}, 3337--3343, \doi{10.1175/2008JAS2770.1}.

\bibitem[{Bolton and Zanna(2019)Bolton, and Zanna}]{Bolton2019}
Bolton, T., and L.~Zanna, 2019: Applications of deep learning to ocean data
  inference and subgrid parameterization. \textit{Journal of Advances in
  Modeling Earth Systems}, \textbf{11~(1)}, 376--399,
  \doi{https://doi.org/10.1029/2018MS001472},
  \urlprefix\url{https://agupubs.onlinelibrary.wiley.com/doi/abs/10.1029/2018MS001472},
  \eprint{https://agupubs.onlinelibrary.wiley.com/doi/pdf/10.1029/2018MS001472}.

\bibitem[{Bouchet et~al.(2011)Bouchet, Laurie,, and Zaboronski}]{Bouchet2011}
Bouchet, F., J.~Laurie, and O.~Zaboronski, 2011: Control and instanton
  trajectories for random transitions in turbulent flows. \textit{Journal of
  Physics: Conference Series}, \textbf{318~(2)}, 022\,041,
  \doi{10.1088/1742-6596/318/2/022041},
  \urlprefix\url{https://doi.org/10.1088%2F1742-6596%2F318%2F2%2F022041}.

\bibitem[{Bouchet et~al.(2014)Bouchet, Laurie,, and
  Zaboronski}]{Bouchet2014_Langevin}
Bouchet, F., J.~Laurie, and O.~Zaboronski, 2014: Langevin dynamics, large
  deviations and instantons for the quasi-geostrophic model and two-dimensional
  euler equations. \textit{Journal of Statistical Physics}, \textbf{156},
  1066--1092, \doi{10.1007/s10955-014-1052-5},
  \urlprefix\url{https://doi.org/10.1007/s10955-014-1052-5}.

\bibitem[{Bouchet et~al.(2019{\natexlab{a}})Bouchet, Rolland,, and
  Simonnet}]{Bouchet2019_nucleations}
Bouchet, F., J.~Rolland, and E.~Simonnet, 2019{\natexlab{a}}: Rare event
  algorithm links transitions in turbulent flows with activated nucleations.
  \textit{Physical Review Letters}, \textbf{122~(7)}, 074\,502,
  \doi{10.1103/PhysRevLett.122.074502}.

\bibitem[{Bouchet et~al.(2019{\natexlab{b}})Bouchet, Rolland,, and
  Wouters}]{Bouchet2019}
Bouchet, F., J.~Rolland, and J.~Wouters, 2019{\natexlab{b}}: Rare event
  sampling methods. \textit{Chaos: An Interdisciplinary Journal of Nonlinear
  Science}, \textbf{29~(8)}, 080\,402, \doi{10.1063/1.5120509}.

\bibitem[{Bowman et~al.(2013)Bowman, Pande,, and
  No{\'e}}]{bowman2013introduction}
Bowman, G.~R., V.~S. Pande, and F.~No{\'e}, 2013: \textit{An introduction to
  Markov state models and their application to long timescale molecular
  simulation}, Vol. 797. Springer Science \& Business Media.

\bibitem[{Carleo and Troyer(2017)Carleo, and Troyer}]{Carleo2017}
Carleo, G., and M.~Troyer, 2017: Solving the quantum many-body problem with
  artificial neural networks. \textit{Science}, \textbf{355~(6325)}, 602--606,
  \doi{10.1126/science.aag2302},
  \urlprefix\url{https://science.sciencemag.org/content/355/6325/602},
  \eprint{https://science.sciencemag.org/content/355/6325/602.full.pdf}.

\bibitem[{Charlton and Polvani(2007)Charlton, and Polvani}]{cp07}
Charlton, A.~J., and L.~M. Polvani, 2007: A new look at stratospheric sudden
  warmings. part i: Climatology and modeling benchmarks. \textit{Journal of
  Climate}, \textbf{20~(3)}, 449--469, \doi{10.1175/JCLI3996.1}.

\bibitem[{Charney and DeVore(1979)Charney, and DeVore}]{Charney1979}
Charney, J.~G., and J.~G. DeVore, 1979: {Multiple Flow Equilibria in the
  Atmosphere and Blocking}. \textit{Journal of the Atmospheric Sciences},
  \textbf{36~(7)}, 1205--1216,
  \doi{10.1175/1520-0469(1979)036<1205:MFEITA>2.0.CO;2},
  \urlprefix\url{https://doi.org/10.1175/1520-0469(1979)036<1205:MFEITA>2.0.CO;2},
  \eprint{https://journals.ametsoc.org/jas/article-pdf/36/7/1205/3420739/1520-0469(1979)036\_1205\_mfeita\_2\_0\_co\_2.pdf}.

\bibitem[{Chattopadhyay et~al.(2021)Chattopadhyay, Mustafa, Hassanzadeh, Bach,,
  and Kashinath}]{Chattopadhyay2021towards}
Chattopadhyay, A., M.~Mustafa, P.~Hassanzadeh, E.~Bach, and K.~Kashinath, 2021:
  Towards physically consistent data-driven weather forecasting: Integrating
  data assimilation with equivariance-preserving spatial transformers in a case
  study with era5. \textit{Geoscientific Model Development Discussions},
  \textbf{2021}, 1--23, \doi{10.5194/gmd-2021-71},
  \urlprefix\url{https://gmd.copernicus.org/preprints/gmd-2021-71/}.

\bibitem[{Chattopadhyay et~al.(2020)Chattopadhyay, Nabizadeh,, and
  Hassanzadeh}]{Chattopadhyay2020}
Chattopadhyay, A., E.~Nabizadeh, and P.~Hassanzadeh, 2020: Analog forecasting
  of extreme-causing weather patterns using deep learning. \textit{Journal of
  Advances in Modeling Earth Systems}, \textbf{12~(2)}, e2019MS001\,958,
  \doi{https://doi.org/10.1029/2019MS001958},
  \urlprefix\url{https://agupubs.onlinelibrary.wiley.com/doi/abs/10.1029/2019MS001958},
  e2019MS001958 10.1029/2019MS001958,
  \eprint{https://agupubs.onlinelibrary.wiley.com/doi/pdf/10.1029/2019MS001958}.

\bibitem[{Chen et~al.(2014)Chen, Giannakis, Herbei,, and Majda}]{Chen2014mcmc}
Chen, N., D.~Giannakis, R.~Herbei, and A.~J. Majda, 2014: An mcmc algorithm for
  parameter estimation in signals with hidden intermittent instability.
  \textit{SIAM/ASA Journal on Uncertainty Quantification}, \textbf{2~(1)},
  647--669, \doi{10.1137/130944977},
  \urlprefix\url{https://doi.org/10.1137/130944977},
  \eprint{https://doi.org/10.1137/130944977}.

\bibitem[{Chen and Majda(2020)Chen, and Majda}]{Chen2020predicting}
Chen, N., and A.~J. Majda, 2020: Predicting observed and hidden extreme events
  in complex nonlinear dynamical systems with partial observations and short
  training time series. \textit{Chaos: An Interdisciplinary Journal of
  Nonlinear Science}, \textbf{30~(3)}, 033\,101, \doi{10.1063/1.5122199},
  \urlprefix\url{https://doi.org/10.1063/1.5122199},
  \eprint{https://doi.org/10.1063/1.5122199}.

\bibitem[{Chodera and Noé(2014)Chodera, and Noé}]{Chodera2014}
Chodera, J.~D., and F.~Noé, 2014: Markov state models of biomolecular
  conformational dynamics. \textit{Current Opinion in Structural Biology},
  \textbf{25}, 135 -- 144, \doi{https://doi.org/10.1016/j.sbi.2014.04.002},
  \urlprefix\url{http://www.sciencedirect.com/science/article/pii/S0959440X14000426},
  theory and simulation / Macromolecular machines.

\bibitem[{Chodera et~al.(2006)Chodera, Swope, Pitera,, and
  Dill}]{Chodera2006longtime}
Chodera, J.~D., W.~C. Swope, J.~W. Pitera, and K.~A. Dill, 2006: Long‐time
  protein folding dynamics from short‐time molecular dynamics simulations.
  \textit{Multiscale Modeling \& Simulation}, \textbf{5~(4)}, 1214--1226,
  \doi{10.1137/06065146X}.

\bibitem[{Christiansen(2000)}]{Christiansen2000}
Christiansen, B., 2000: Chaos, quasiperiodicity, and interannual variability:
  Studies of a stratospheric vacillation model. \textit{Journal of the
  Atmospheric Sciences}, \textbf{57~(18)}, 3161--3173,
  \doi{10.1175/1520-0469(2000)057<3161:CQAIVS>2.0.CO;2}.

\bibitem[{Crommelin(2003)}]{Crommelin2003}
Crommelin, D.~T., 2003: Regime transitions and heteroclinic connections in a
  barotropic atmosphere. \textit{Journal of the Atmospheric Sciences},
  \textbf{60~(2)}, 229 -- 246,
  \doi{10.1175/1520-0469(2003)060<0229:RTAHCI>2.0.CO;2},
  \urlprefix\url{https://journals.ametsoc.org/view/journals/atsc/60/2/1520-0469_2003_060_0229_rtahci_2.0.co_2.xml}.

\bibitem[{DelSole and Farrell(1995)DelSole, and Farrell}]{Delsole1995}
DelSole, T., and B.~F. Farrell, 1995: A stochastically excited linear system as
  a model for quasigeostrophic turbulence: Analytic results for one- and
  two-layer fluids. \textit{Journal of the Atmospheric Sciences},
  \textbf{52~(14)}, 2531--2547,
  \doi{10.1175/1520-0469(1995)052<2531:ASELSA>2.0.CO;2}.

\bibitem[{Dematteis et~al.(2019)Dematteis, Grafke, Onorato,, and
  Vanden-Eijnden}]{Dematteis2019experimental}
Dematteis, G., T.~Grafke, M.~Onorato, and E.~Vanden-Eijnden, 2019: Experimental
  evidence of hydrodynamic instantons: The universal route to rogue waves.
  \textit{Phys. Rev. X}, \textbf{9}, 041\,057, \doi{10.1103/PhysRevX.9.041057},
  \urlprefix\url{https://link.aps.org/doi/10.1103/PhysRevX.9.041057}.

\bibitem[{Dematteis et~al.(2018)Dematteis, Grafke,, and Vanden-Eijnden}]{rogue}
Dematteis, G., T.~Grafke, and E.~Vanden-Eijnden, 2018: Rogue waves and large
  deviations in deep sea. \textit{Proceedings of the National Academy of
  Sciences}, \textbf{115~(5)}, 855--860, \doi{10.1073/pnas.1710670115}.

\bibitem[{Durrett(2013)}]{Durrett}
Durrett, R., 2013: \textit{Probability: Theory and Examples}. Cambridge
  University Press.

\bibitem[{E et~al.(2019)E, Li,, and Vanden-Eijnden}]{weinan2019applied}
E, W., T.~Li, and E.~Vanden-Eijnden, 2019: \textit{Applied stochastic
  analysis}, Vol. 199. American Mathematical Soc.

\bibitem[{E. and Vanden-Eijnden(2006)E., and Vanden-Eijnden}]{Weinan2006}
E., W., and E.~Vanden-Eijnden, 2006: Towards a {Theory} of {Transition}
  {Paths}. \textit{Journal of Statistical Physics}, \textbf{123~(3)}, 503,
  \doi{10.1007/s10955-005-9003-9},
  \urlprefix\url{https://doi.org/10.1007/s10955-005-9003-9}.

\bibitem[{Esler and Mester(2019)Esler, and Mester}]{Esler2019}
Esler, J.~G., and M.~Mester, 2019: Noise-induced vortex-splitting stratospheric
  sudden warmings. \textit{Quarterly Journal of the Royal Meteorological
  Society}, \textbf{145~(719)}, 476--494,
  \doi{https://doi.org/10.1002/qj.3443},
  \urlprefix\url{https://rmets.onlinelibrary.wiley.com/doi/abs/10.1002/qj.3443},
  \eprint{https://rmets.onlinelibrary.wiley.com/doi/pdf/10.1002/qj.3443}.

\bibitem[{Farazmand and Sapsis(2017)Farazmand, and
  Sapsis}]{Farazmand2017variational}
Farazmand, M., and T.~P. Sapsis, 2017: A variational approach to probing
  extreme events in turbulent dynamical systems. \textit{Science Advances},
  \textbf{3~(9)}, \doi{10.1126/sciadv.1701533},
  \urlprefix\url{https://advances.sciencemag.org/content/3/9/e1701533},
  \eprint{https://advances.sciencemag.org/content/3/9/e1701533.full.pdf}.

\bibitem[{Finkel et~al.(2020)Finkel, Abbot,, and Weare}]{Finkel2020}
Finkel, J., D.~S. Abbot, and J.~Weare, 2020: {Path Properties of Atmospheric
  Transitions: Illustration with a Low-Order Sudden Stratospheric Warming
  Model}. \textit{Journal of the Atmospheric Sciences}, \textbf{77~(7)},
  2327--2347, \doi{10.1175/JAS-D-19-0278.1},
  \urlprefix\url{https://doi.org/10.1175/JAS-D-19-0278.1},
  \eprint{https://journals.ametsoc.org/jas/article-pdf/77/7/2327/4958190/jasd190278.pdf}.

\bibitem[{Fitzsimmons and Pitman(1999)Fitzsimmons, and
  Pitman}]{Fitzsimmons1999kac}
Fitzsimmons, P., and J.~Pitman, 1999: Kac’s moment formula and the
  feynman–kac formula for additive functionals of a markov process.
  \textit{Stochastic Processes and their Applications}, \textbf{79~(1)},
  117--134, \doi{https://doi.org/10.1016/S0304-4149(98)00081-7},
  \urlprefix\url{https://www.sciencedirect.com/science/article/pii/S0304414998000817}.

\bibitem[{Frank and Fischer(2008)Frank, and Fischer}]{Noe2008}
Frank, N., and S.~Fischer, 2008: Transition networks for modeling the kinetics
  of conformational change in macromolecules. \textit{Current Opininion in
  Structural Biology}, \textbf{18}, 154--163, \doi{10.1016/j.sbi.2008.01.008}.

\bibitem[{Franzke and Majda(2006)Franzke, and Majda}]{franzke_majda2006}
Franzke, C., and A.~J. Majda, 2006: Low-order stochastic mode reduction for a
  prototype atmospheric gcm. \textit{Journal of the Atmospheric Sciences},
  \textbf{63~(2)}, 457--479, \doi{10.1175/JAS3633.1}.

\bibitem[{Giannakis et~al.(2018)Giannakis, Kolchinskaya, Krasnov,, and
  Schumacher}]{giannakis_kolchinskaya_krasnov_schumacher_2018}
Giannakis, D., A.~Kolchinskaya, D.~Krasnov, and J.~Schumacher, 2018: Koopman
  analysis of the long-term evolution in a turbulent convection cell.
  \textit{Journal of Fluid Mechanics}, \textbf{847}, 735–767,
  \doi{10.1017/jfm.2018.297}.

\bibitem[{Giannakis and Majda(2012)Giannakis, and Majda}]{nlsa}
Giannakis, D., and A.~J. Majda, 2012: Nonlinear laplacian spectral analysis for
  time series with intermittency and low-frequency variability.
  \textit{Proceedings of the National Academy of Sciences}, \textbf{109~(7)},
  2222--2227, \doi{10.1073/pnas.1118984109},
  \eprint{https://www.pnas.org/content/109/7/2222.full.pdf}.

\bibitem[{Gottwald et~al.(2016)Gottwald, Crommelin,, and
  Franzke}]{gottwald2016stochastic}
Gottwald, G.~A., D.~T. Crommelin, and C.~L.~E. Franzke, 2016: Stochastic
  climate theory. \eprint{1612.07474}.

\bibitem[{Han et~al.(2018)Han, Jentzen,, and E}]{Han2018}
Han, J., A.~Jentzen, and W.~E, 2018: Solving high-dimensional partial
  differential equations using deep learning. \textit{Proceedings of the
  National Academy of Sciences}, \textbf{115~(34)}, 8505--8510,
  \doi{10.1073/pnas.1718942115},
  \urlprefix\url{https://www.pnas.org/content/115/34/8505},
  \eprint{https://www.pnas.org/content/115/34/8505.full.pdf}.

\bibitem[{Hasselmann(1976)}]{hasselman}
Hasselmann, K., 1976: Stochastic climate models part i. theory.
  \textit{Tellus}, \textbf{28~(6)}, 473--485,
  \doi{10.3402/tellusa.v28i6.11316}.

\bibitem[{Helfmann et~al.(2021)Helfmann, Heitzig, Koltai, Kurths,, and
  Schütte}]{Helfmann2021statistical}
Helfmann, L., J.~Heitzig, P.~Koltai, J.~Kurths, and C.~Schütte, 2021:
  Statistical analysis of tipping pathways in agent-based models.
  \eprint{2103.02883}.

\bibitem[{Hoffman et~al.(2006)Hoffman, Henderson, Leidner, Grassotti,, and
  Nehrkorn}]{Hoffman2006response}
Hoffman, R.~N., J.~M. Henderson, S.~M. Leidner, C.~Grassotti, and T.~Nehrkorn,
  2006: The response of damaging winds of a simulated tropical cyclone to
  finite-amplitude perturbations of different variables. \textit{Journal of the
  Atmospheric Sciences}, \textbf{63~(7)}, 1924 -- 1937,
  \doi{10.1175/JAS3720.1},
  \urlprefix\url{https://journals.ametsoc.org/view/journals/atsc/63/7/jas3720.1.xml}.

\bibitem[{Holton and Mass(1976)Holton, and Mass}]{holton_mass}
Holton, J.~R., and C.~Mass, 1976: Stratospheric vacillation cycles.
  \textit{Journal of the Atmospheric Sciences}, \textbf{33~(11)}, 2218--2225,
  \doi{10.1175/1520-0469(1976)033<2218:SVC>2.0.CO;2}.

\bibitem[{Jayachandran et~al.(2006)Jayachandran, Vishal,, and
  Pande}]{Jayachandran2006folding}
Jayachandran, G., V.~Vishal, and V.~S. Pande, 2006: Using massively parallel
  simulation and markovian models to study protein folding: Examining the
  dynamics of the villin headpiece. \textit{The Journal of Chemical Physics},
  \textbf{124~(16)}, 164\,902, \doi{10.1063/1.2186317},
  \urlprefix\url{https://doi.org/10.1063/1.2186317},
  \eprint{https://doi.org/10.1063/1.2186317}.

\bibitem[{Karatzas and Shreve(1998)Karatzas, and Shreve}]{Karatzas}
Karatzas, I., and S.~E. Shreve, 1998: \textit{Brownian Motion and Stochastic
  Calculus}. Springer.

\bibitem[{Kashinath et~al.(2021)}]{Kashinath2021physics}
Kashinath, K., and Coauthors, 2021: Physics-informed machine learning: case
  studies for weather and climate modelling. \textit{Philosophical Transactions
  of the Royal Society A: Mathematical, Physical and Engineering Sciences},
  \textbf{379~(2194)}, 20200\,093, \doi{10.1098/rsta.2020.0093},
  \urlprefix\url{https://royalsocietypublishing.org/doi/abs/10.1098/rsta.2020.0093}.

\bibitem[{Khoo et~al.(2018)Khoo, Lu,, and Ying}]{Khoo2018}
Khoo, Y., J.~Lu, and L.~Ying, 2018: Solving for high-dimensional committor
  functions using artificial neural networks. \textit{Research in the
  Mathematical Sciences}, \textbf{6}, \doi{10.1007/s40687-018-0160-2},
  \urlprefix\url{https://doi.org/10.1007/s40687-018-0160-2}.

\bibitem[{Legras and Ghil(1985)Legras, and Ghil}]{Legras1985}
Legras, B., and M.~Ghil, 1985: Persistent anomalies, blocking and variations in
  atmospheric predictability. \textit{Journal of Atmospheric Sciences},
  \textbf{42~(5)}, 433 -- 471,
  \doi{10.1175/1520-0469(1985)042<0433:PABAVI>2.0.CO;2},
  \urlprefix\url{https://journals.ametsoc.org/view/journals/atsc/42/5/1520-0469_1985_042_0433_pabavi_2_0_co_2.xml}.

\bibitem[{Li et~al.(2020)Li, Khoo, Ren,, and Ying}]{Li2020}
Li, H., Y.~Khoo, Y.~Ren, and L.~Ying, 2020: Solving for high dimensional
  committor functions using neural network with online approximation to
  derivatives. \eprint{2012.06727}.

\bibitem[{Li et~al.(2019)Li, Lin,, and Ren}]{Li2019computing}
Li, Q., B.~Lin, and W.~Ren, 2019: Computing committor functions for the study
  of rare events using deep learning. \textit{The Journal of Chemical Physics},
  \textbf{151~(5)}, 054\,112, \doi{10.1063/1.5110439},
  \urlprefix\url{https://doi.org/10.1063/1.5110439}.

\bibitem[{Lin and Lu(2021)Lin, and Lu}]{Lin2021koopman}
Lin, K.~K., and F.~Lu, 2021: Data-driven model reduction, wiener projections,
  and the koopman-mori-zwanzig formalism. \textit{Journal of Computational
  Physics}, \textbf{424}, 109\,864,
  \doi{https://doi.org/10.1016/j.jcp.2020.109864},
  \urlprefix\url{https://www.sciencedirect.com/science/article/pii/S0021999120306380}.

\bibitem[{Lorenz(1963)}]{Lorenz1963}
Lorenz, E.~N., 1963: Deterministic nonperiodic flow. \textit{Journal of
  Atmospheric Sciences}, \textbf{20~(2)}, 130 -- 141,
  \doi{10.1175/1520-0469(1963)020<0130:DNF>2.0.CO;2},
  \urlprefix\url{https://journals.ametsoc.org/view/journals/atsc/20/2/1520-0469_1963}.

\bibitem[{Lorpaiboon et~al.(2020)Lorpaiboon, Thiede, Webber, Weare,, and
  Dinner}]{Lorpaiboon2020ivac}
Lorpaiboon, C., E.~H. Thiede, R.~J. Webber, J.~Weare, and A.~R. Dinner, 2020:
  Integrated variational approach to conformational dynamics: A robust strategy
  for identifying eigenfunctions of dynamical operators. \textit{The Journal of
  Physical Chemistry B}, \textbf{124~(42)}, 9354--9364,
  \doi{10.1021/acs.jpcb.0c06477},
  \urlprefix\url{https://doi.org/10.1021/acs.jpcb.0c06477}.

\bibitem[{Lucarini and Gritsun(2020)Lucarini, and Gritsun}]{Lucarini2020new}
Lucarini, V., and A.~Gritsun, 2020: A new mathematical framework for
  atmospheric blocking events. \textit{Climate Dynamics}, \textbf{54~(1)},
  575--598.

\bibitem[{Lucente et~al.(2019)Lucente, Duffner, Herbert, Rolland,, and
  Bouchet}]{Lucente2019}
Lucente, D., S.~Duffner, C.~Herbert, J.~Rolland, and F.~Bouchet, 2019: {Machine
  learning of committor functions for predicting high impact climate events}.
  \textit{{Climate Informatics}}, Paris, France,
  \urlprefix\url{https://hal.archives-ouvertes.fr/hal-02322370}.

\bibitem[{{Maiocchi} et~al.(2020){Maiocchi}, {Lucarini}, {Gritsun},, and
  {Pavliotis}}]{Maiocchi2020unstable}
{Maiocchi}, C.~C., V.~{Lucarini}, A.~{Gritsun}, and G.~{Pavliotis}, 2020:
  {Unstable Periodic Orbits Sampling in Climate Models}. \textit{EGU General
  Assembly Conference Abstracts}, 18823, EGU General Assembly Conference
  Abstracts.

\bibitem[{Majda and Qi(2018)Majda, and Qi}]{Majda2018strategies}
Majda, A.~J., and D.~Qi, 2018: Strategies for reduced-order models for
  predicting the statistical responses and uncertainty quantification in
  complex turbulent dynamical systems. \textit{SIAM Review}, \textbf{60~(3)},
  491--549, \doi{10.1137/16M1104664},
  \urlprefix\url{https://doi.org/10.1137/16M1104664},
  \eprint{https://doi.org/10.1137/16M1104664}.

\bibitem[{Majda et~al.(2001)Majda, Timofeyev,, and
  Vanden~Eijnden}]{Majda2001stochastic}
Majda, A.~J., I.~Timofeyev, and E.~Vanden~Eijnden, 2001: A mathematical
  framework for stochastic climate models. \textit{Communications on Pure and
  Applied Mathematics}, \textbf{54~(8)}, 891--974,
  \doi{https://doi.org/10.1002/cpa.1014},
  \urlprefix\url{https://onlinelibrary.wiley.com/doi/abs/10.1002/cpa.1014},
  \eprint{https://onlinelibrary.wiley.com/doi/pdf/10.1002/cpa.1014}.

\bibitem[{Mardt et~al.(2018)Mardt, Pasuali, Wu,, and No\'e}]{Mardt2018vampnets}
Mardt, A., L.~Pasuali, H.~Wu, and F.~No\'e, 2018: Vampnets for deep learning of
  molecular kinetics. \textit{Nature Communications}, \textbf{9},
  \doi{10.1038/s41467-017-02388-1},
  \urlprefix\url{https://doi.org/10.1038/s41467-017-02388-1}.

\bibitem[{Matsuno(1971)}]{Matsuno1971}
Matsuno, T., 1971: {A Dynamical Model of the Stratospheric Sudden Warming}.
  \textit{Journal of the Atmospheric Sciences}, \textbf{28~(8)}, 1479--1494,
  \doi{10.1175/1520-0469(1971)028<1479:ADMOTS>2.0.CO;2},
  \urlprefix\url{https://doi.org/10.1175/1520-0469(1971)028<1479:ADMOTS>2.0.CO;2},
  \eprint{https://journals.ametsoc.org/jas/article-pdf/28/8/1479/3417422/1520-0469(1971)028\_1479\_admots\_2\_0\_co\_2.pdf}.

\bibitem[{Metzner et~al.(2006)Metzner, Schutte,, and
  Vanden-Eijnden}]{tpt_simple_examples}
Metzner, P., C.~Schutte, and E.~Vanden-Eijnden, 2006: Illustration of
  transition path theory on a collection of simple examples. \textit{The
  Journal of Chemical Physics}, \textbf{125~(8)}, 1--2,
  \doi{10.1063/1.2335447}.

\bibitem[{Metzner et~al.(2009)Metzner, Schutte,, and Vanden-Eijnden}]{tpt_mjp}
Metzner, P., C.~Schutte, and E.~Vanden-Eijnden, 2009: Transition path theory
  for markov jump processes. \textit{Multiscale Modeling and Simulation},
  \textbf{7~(3)}, 1192--1219, \doi{10.1137/070699500}.

\bibitem[{Miron et~al.(2021)Miron, Beron-Vera, Helfmann,, and
  Koltai}]{Miron2020}
Miron, P., F.~Beron-Vera, L.~Helfmann, and P.~Koltai, 2021: Transition paths of
  marine debris and the stability of the garbage patches. \textit{Chaos: An
  Interdisciplinary Journal of Nonlinear Science}, accepted for publication.

\bibitem[{Mohamad and Sapsis(2018)Mohamad, and Sapsis}]{Mohamad2018sequential}
Mohamad, M.~A., and T.~P. Sapsis, 2018: Sequential sampling strategy for
  extreme event statistics in nonlinear dynamical systems. \textit{Proceedings
  of the National Academy of Sciences}, \textbf{115~(44)}, 11\,138--11\,143,
  \doi{10.1073/pnas.1813263115},
  \urlprefix\url{https://www.pnas.org/content/115/44/11138},
  \eprint{https://www.pnas.org/content/115/44/11138.full.pdf}.

\bibitem[{Ngwira et~al.(2013)}]{Ngwira2013}
Ngwira, C.~M., and Coauthors, 2013: Simulation of the 23 july 2012 extreme
  space weather event: What if this extremely rare cme was earth directed?
  \textit{Space Weather}, \textbf{11~(12)}, 671--679,
  \doi{https://doi.org/10.1002/2013SW000990},
  \urlprefix\url{https://agupubs.onlinelibrary.wiley.com/doi/abs/10.1002/2013SW000990},
  \eprint{https://agupubs.onlinelibrary.wiley.com/doi/pdf/10.1002/2013SW000990}.

\bibitem[{No{\'e} et~al.(2009)No{\'e}, Sch{\"u}tte, Vanden-Eijnden, Reich,, and
  Weikl}]{Noe2009}
No{\'e}, F., C.~Sch{\"u}tte, E.~Vanden-Eijnden, L.~Reich, and T.~R. Weikl,
  2009: Constructing the equilibrium ensemble of folding pathways from short
  off-equilibrium simulations. \textit{Proceedings of the National Academy of
  Sciences}, \textbf{106~(45)}, 19\,011--19\,016,
  \doi{10.1073/pnas.0905466106},
  \urlprefix\url{https://www.pnas.org/content/106/45/19011},
  \eprint{https://www.pnas.org/content/106/45/19011.full.pdf}.

\bibitem[{Noé and Clementi(2017)Noé, and Clementi}]{Noe2017cv}
Noé, F., and C.~Clementi, 2017: Collective variables for the study of
  long-time kinetics from molecular trajectories: theory and methods.
  \textit{Current Opinion in Structural Biology}, \textbf{43}, 141--147,
  \doi{https://doi.org/10.1016/j.sbi.2017.02.006},
  \urlprefix\url{https://www.sciencedirect.com/science/article/pii/S0959440X17300301},
  theory and simulation • Macromolecular assemblies.

\bibitem[{Oksendal(2003)}]{Oksendal}
Oksendal, B., 2003: \textit{Stochastic Differential Equations: An Introduction
  with Applications}. Springer.

\bibitem[{Pande et~al.(2010)Pande, Beauchamp,, and Bowman}]{Pande2010}
Pande, V.~S., K.~Beauchamp, and G.~R. Bowman, 2010: Everything you wanted to
  know about markov state models but were afraid to ask. \textit{Methods},
  \textbf{52~(1)}, 99–105,
  \urlprefix\url{https://doi.org/10.1016/j.ymeth.2010.06.002}.

\bibitem[{Pedregosa et~al.(2011)}]{scikit-learn}
Pedregosa, F., and Coauthors, 2011: Scikit-learn: Machine learning in {P}ython.
  \textit{Journal of Machine Learning Research}, \textbf{12}, 2825--2830.

\bibitem[{Plotkin et~al.(2019)Plotkin, Webber, O'Neill, Weare,, and
  Abbot}]{Plotkin2019}
Plotkin, D.~A., R.~J. Webber, M.~E. O'Neill, J.~Weare, and D.~S. Abbot, 2019:
  Maximizing simulated tropical cyclone intensity with action minimization.
  \textit{Journal of Advances in Modeling Earth Systems}, \textbf{11~(4)},
  863--891, \doi{10.1029/2018MS001419}.

\bibitem[{{Porta Mana} and Zanna(2014){Porta Mana}, and
  Zanna}]{Portamana2014stochastic}
{Porta Mana}, P., and L.~Zanna, 2014: Toward a stochastic parameterization of
  ocean mesoscale eddies. \textit{Ocean Modelling}, \textbf{79}, 1--20,
  \doi{https://doi.org/10.1016/j.ocemod.2014.04.002},
  \urlprefix\url{https://www.sciencedirect.com/science/article/pii/S1463500314000420}.

\bibitem[{Ragone and Bouchet(2020)Ragone, and Bouchet}]{Ragone2020averaged}
Ragone, F., and F.~Bouchet, 2020: Computation of extreme values of time
  averaged observables in climate models with large deviation techniques.
  \textit{Journal of Statistical Physics}, \textbf{179~(5)}, 1637--1665,
  \doi{10.1007/s10955-019-02429-7},
  \urlprefix\url{https://doi.org/10.1007/s10955-019-02429-7}.

\bibitem[{Ragone et~al.(2018)Ragone, Wouters,, and Bouchet}]{Ragone24}
Ragone, F., J.~Wouters, and F.~Bouchet, 2018: Computation of extreme heat waves
  in climate models using a large deviation algorithm. \textit{Proceedings of
  the National Academy of Sciences}, \textbf{115~(1)}, 24--29,
  \doi{10.1073/pnas.1712645115},
  \eprint{https://www.pnas.org/content/115/1/24.full.pdf}.

\bibitem[{Raissi et~al.(2019)Raissi, Perdikaris,, and
  Karniadakis}]{Raissi2019pinn}
Raissi, M., P.~Perdikaris, and G.~Karniadakis, 2019: Physics-informed neural
  networks: A deep learning framework for solving forward and inverse problems
  involving nonlinear partial differential equations. \textit{Journal of
  Computational Physics}, \textbf{378}, 686--707,
  \doi{https://doi.org/10.1016/j.jcp.2018.10.045},
  \urlprefix\url{https://www.sciencedirect.com/science/article/pii/S0021999118307125}.

\bibitem[{Rotskoff and Vanden-Eijnden(2020)Rotskoff, and
  Vanden-Eijnden}]{rotskoff2020learning}
Rotskoff, G.~M., and E.~Vanden-Eijnden, 2020: Learning with rare data: Using
  active importance sampling to optimize objectives dominated by rare events.
  \eprint{2008.06334}.

\bibitem[{Ruzmaikin et~al.(2003)Ruzmaikin, Lawrence,, and Cadavid}]{ruz}
Ruzmaikin, A., J.~Lawrence, and C.~Cadavid, 2003: A simple model of
  stratospheric dynamics including solar variability. \textit{Journal of
  Climate}, \textbf{16}, 1593--1600, \doi{10.1175/2007JCLI2119.1}.

\bibitem[{Sabeerali et~al.(2017)Sabeerali, Ajayamohan, Giannakis,, and
  Majda}]{Sabeerali2017}
Sabeerali, C.~T., R.~S. Ajayamohan, D.~Giannakis, and A.~J. Majda, 2017:
  Extraction and prediction of indices for monsoon intraseasonal oscillations:
  an approach based on nonlinear laplacian spectral analysis. \textit{Climate
  Dynamics}, \textbf{49~(9)}, 3031--3050, \doi{10.1007/s00382-016-3491-y}.

\bibitem[{Sapsis(2021)}]{Sapsis2021statistics}
Sapsis, T.~P., 2021: Statistics of extreme events in fluid flows and waves.
  \textit{Annual Review of Fluid Mechanics}, \textbf{53~(1)}, 85--111,
  \doi{10.1146/annurev-fluid-030420-032810},
  \urlprefix\url{https://doi.org/10.1146/annurev-fluid-030420-032810},
  \eprint{https://doi.org/10.1146/annurev-fluid-030420-032810}.

\bibitem[{Schaller et~al.(2018)Schaller, Sillmann, Anstey, Fischer, Grams,, and
  Russo}]{Schaller2018}
Schaller, N., J.~Sillmann, J.~Anstey, E.~M. Fischer, C.~M. Grams, and S.~Russo,
  2018: Influence of blocking on northern european and western russian
  heatwaves in large climate model ensembles. \textit{Environmental Research
  Letters}, \textbf{13~(5)}, 054\,015, \doi{10.1088/1748-9326/aaba55},
  \urlprefix\url{https://doi.org/10.1088%2F1748-9326%2Faaba55}.

\bibitem[{Simonnet et~al.(2020)Simonnet, Rolland,, and
  Bouchet}]{simonnet2020multistability}
Simonnet, E., J.~Rolland, and F.~Bouchet, 2020: Multistability and rare
  spontaneous transitions between climate and jet configurations in a
  barotropic model of the jovian mid-latitude troposphere. \eprint{2009.09913}.

\bibitem[{Sjoberg and Birner(2014)Sjoberg, and Birner}]{Sjoberg2014flux}
Sjoberg, J.~P., and T.~Birner, 2014: Stratospheric wave–mean flow feedbacks
  and sudden stratospheric warmings in a simple model forced by upward wave
  activity flux. \textit{Journal of the Atmospheric Sciences},
  \textbf{71~(11)}, 4055 -- 4071, \doi{10.1175/JAS-D-14-0113.1},
  \urlprefix\url{https://journals.ametsoc.org/view/journals/atsc/71/11/jas-d-14-0113.1.xml}.

\bibitem[{Strahan et~al.(2021)Strahan, Antoszewski, Lorpaiboon, Vani, Weare,,
  and Dinner}]{Strahan2020}
Strahan, J., A.~Antoszewski, C.~Lorpaiboon, B.~P. Vani, J.~Weare, and A.~R.
  Dinner, 2021: Long-time-scale predictions from short-trajectory data: A
  benchmark analysis of the trp-cage miniprotein. \textit{Journal of Chemical
  Theory and Computation}, \textbf{17~(5)}, 2948--2963,
  \doi{10.1021/acs.jctc.0c00933},
  \urlprefix\url{https://doi.org/10.1021/acs.jctc.0c00933}, pMID: 33908762,
  \eprint{https://doi.org/10.1021/acs.jctc.0c00933}.

\bibitem[{Tantet et~al.(2015)Tantet, van~der Burgt,, and Dijkstra}]{Tantet2015}
Tantet, A., F.~R. van~der Burgt, and H.~A. Dijkstra, 2015: An early warning
  indicator for atmospheric blocking events using transfer operators.
  \textit{Chaos: An Interdisciplinary Journal of Nonlinear Science},
  \textbf{25~(3)}, 036\,406, \doi{10.1063/1.4908174},
  \urlprefix\url{https://doi.org/10.1063/1.4908174},
  \eprint{https://doi.org/10.1063/1.4908174}.

\bibitem[{Thiede et~al.(2019)Thiede, Giannakis, Dinner,, and Weare}]{dga}
Thiede, E., D.~Giannakis, A.~R. Dinner, and J.~Weare, 2019: Approximation of
  dynamical quantities using trajectory data. \textit{arXiv:1810.01841
  [physics.data-an]}, 1--24, \doi{1810.01841}.

\bibitem[{Tibshirani(1996)}]{Tibshirani1996}
Tibshirani, R., 1996: Regression shrinkage and selection via the lasso.
  \textit{Journal of the Royal Statistical Society: Series B (Methodological)},
  \textbf{58~(1)}, 267--288,
  \doi{https://doi.org/10.1111/j.2517-6161.1996.tb02080.x},
  \urlprefix\url{https://rss.onlinelibrary.wiley.com/doi/abs/10.1111/j.2517-6161.1996.tb02080.x},
  \eprint{https://rss.onlinelibrary.wiley.com/doi/pdf/10.1111/j.2517-6161.1996.tb02080.x}.

\bibitem[{Timmermann et~al.(2003)Timmermann, Jin,, and
  Abshagen}]{Timmerman2003}
Timmermann, A., F.-F. Jin, and J.~Abshagen, 2003: A nonlinear theory for el
  ni\~{n}o bursting. \textit{Journal of the Atmospheric Sciences},
  \textbf{60~(1)}, 152 -- 165,
  \doi{10.1175/1520-0469(2003)060<0152:ANTFEN>2.0.CO;2},
  \urlprefix\url{https://journals.ametsoc.org/view/journals/atsc/60/1/1520-0469_2003}.

\bibitem[{Vanden-Eijnden and E(2010)Vanden-Eijnden, and E}]{pathfinding}
Vanden-Eijnden, E., and W.~E, 2010: Transition-path theory and path-finding
  algorithms for the study of rare events. \textit{Annual Review of Physical
  Chemistry}, \textbf{61~(1)}, 391--420,
  \doi{10.1146/annurev.physchem.040808.090412}.

\bibitem[{Vanden-Eijnden and Weare(2013)Vanden-Eijnden, and
  Weare}]{VandenEijnden2013}
Vanden-Eijnden, E., and J.~Weare, 2013: Data assimilation in the low noise
  regime with application to the kuroshio. \textit{Monthly Weather Review},
  \textbf{141~(6)}, 1822--1841, \doi{10.1175/MWR-D-12-00060.1}.

\bibitem[{Vitart and Robertson(2018)Vitart, and Robertson}]{Vitart2018}
Vitart, F., and A.~W. Robertson, 2018: The sub-seasonal to seasonal prediction
  project (s2s) and the prediction of extreme events. \textit{npj Climate and
  Atmospheric Science}, \textbf{1},
  \urlprefix\url{https://doi.org/10.1038/s41612-018-0013-0}.

\bibitem[{Wan et~al.(2018)Wan, Vlachas, Koumoutsakos,, and
  Sapsis}]{Wan2018data-assisted}
Wan, Z.~Y., P.~Vlachas, P.~Koumoutsakos, and T.~Sapsis, 2018: Data-assisted
  reduced-order modeling of extreme events in complex dynamical systems.
  \textit{PLOS ONE}, \textbf{13~(5)}, 1--22,
  \doi{10.1371/journal.pone.0197704},
  \urlprefix\url{https://doi.org/10.1371/journal.pone.0197704}.

\bibitem[{Weare(2009)}]{Weare2009}
Weare, J., 2009: Particle filtering with path sampling and an application to a
  bimodal ocean current model. \textit{Journal of Computational Physics},
  \textbf{228~(12)}, 4312 -- 4331,
  \doi{https://doi.org/10.1016/j.jcp.2009.02.033}.

\bibitem[{Webber et~al.(2019)Webber, Plotkin, O’Neill, Abbot,, and
  Weare}]{webber}
Webber, R.~J., D.~A. Plotkin, M.~E. O’Neill, D.~S. Abbot, and J.~Weare, 2019:
  Practical rare event sampling for extreme mesoscale weather. \textit{Chaos},
  \textbf{29~(5)}, 053\,109, \doi{10.1063/1.5081461}.

\bibitem[{Yasuda et~al.(2017)Yasuda, Bouchet,, and Venaille}]{bouchet}
Yasuda, Y., F.~Bouchet, and A.~Venaille, 2017: A new interpretation of
  vortex-split sudden stratospheric warmings in terms of equilibrium
  statistical mechanics. \textit{Journal of the Atmospheric Sciences},
  \textbf{74~(12)}, 3915--3936, \doi{10.1175/JAS-D-17-0045.1}.

\bibitem[{Yoden(1987{\natexlab{a}})}]{Yoden1987_bif}
Yoden, S., 1987{\natexlab{a}}: Bifurcation properties of a stratospheric
  vacillation model. \textit{Journal of the Atmospheric Sciences},
  \textbf{44~(13)}, 1723--1733,
  \doi{10.1175/1520-0469(1987)044<1723:BPOASV>2.0.CO;2}.

\bibitem[{Yoden(1987{\natexlab{b}})}]{Yoden1987_dyn}
Yoden, S., 1987{\natexlab{b}}: {Dynamical Aspects of Stratospheric Vacillations
  in a Highly Truncated Model}. \textit{Journal of the Atmospheric Sciences},
  \textbf{44~(24)}, 3683--3695,
  \doi{10.1175/1520-0469(1987)044<3683:DAOSVI>2.0.CO;2},
  \urlprefix\url{https://doi.org/10.1175/1520-0469(1987)044<3683:DAOSVI>2.0.CO;2}.

\bibitem[{Zhang and Sippel(2009)Zhang, and Sippel}]{Zhang2009}
Zhang, F., and J.~A. Sippel, 2009: Effects of moist convection on hurricane
  predictability. \textit{Journal of the Atmospheric Sciences},
  \textbf{66~(7)}, 1944 -- 1961, \doi{10.1175/2009JAS2824.1},
  \urlprefix\url{https://journals.ametsoc.org/view/journals/atsc/66/7/2009jas2824.1.xml}.

\bibitem[{Zwanzig(2001)}]{zwanzig}
Zwanzig, R., 2001: \textit{Nonequilibrium statistical mechanics}. Oxford
  University Press.

\end{thebibliography}

%%%%%%%%%%%%%%%%%%%%%%%%%%%%%%%%%%%%%%%%%%%%%%%%%%%%%%%%%%%%%%%%%%%%%
% TABLES
%%%%%%%%%%%%%%%%%%%%%%%%%%%%%%%%%%%%%%%%%%%%%%%%%%%%%%%%%%%%%%%%%%%%%
%% Enter tables at the end of the document, before figures.
%%
%
%\begin{table}[t]
%\caption{This is a sample table caption and table layout.  Enter as many tables as
%  necessary at the end of your manuscript. Table from Lorenz (1963).}\label{t1}
%\begin{center}
%\begin{tabular}{ccccrrcrc}
%\hline\hline
%$N$ & $X$ & $Y$ & $Z$\\
%\hline
% 0000 & 0000 & 0010 & 0000 \\
% 0005 & 0004 & 0012 & 0000 \\
% 0010 & 0009 & 0020 & 0000 \\
% 0015 & 0016 & 0036 & 0002 \\
% 0020 & 0030 & 0066 & 0007 \\
% 0025 & 0054 & 0115 & 0024 \\
%\hline
%\end{tabular}
%\end{center}
%\end{table}

%%%%%%%%%%%%%%%%%%%%%%%%%%%%%%%%%%%%%%%%%%%%%%%%%%%%%%%%%%%%%%%%%%%%%
% FIGURES
%%%%%%%%%%%%%%%%%%%%%%%%%%%%%%%%%%%%%%%%%%%%%%%%%%%%%%%%%%%%%%%%%%%%%
%% Enter figures at the end of the document, after tables.
%%
%
%\begin{figure}[t]
%  \noindent\includegraphics[width=19pc,angle=0]{figure01.pdf}\\
%  \caption{Enter the caption for your figure here.  Repeat as
%  necessary for each of your figures. Fig. from \protect\cite{Knutti2008}.}\label{f1}
%\end{figure}

\end{document}